\documentclass{aa}  

\usepackage{color}
\usepackage[final]{changes}
\usepackage{graphicx}
\usepackage{txfonts}
\usepackage[]{hyperref}
\usepackage{amstext}

\begin{document} 

\title{Substructure in the stellar halo near the Sun. I. Data-driven clustering in integrals-of-motion space\thanks{Tables~\ref{table:cluster_description} and \ref{table:catalogue} are only available in electronic form
at the CDS via anonymous ftp to cdsarc.u-strasbg.fr (130.79.128.5)
or via \url{http://cdsweb.u-strasbg.fr/cgi-bin/qcat?J/A+A/}}}

\titlerunning{Clustering in integrals-of-motion space}
\authorrunning{S. S. L\"{o}vdal et al.}

\author{S. Sofie L\"{o}vdal
    \inst{1}, Tom\'{a}s Ruiz-Lara\inst{1}, 
    Helmer H. Koppelman\inst{2}, Tadafumi Matsuno\inst{1}, Emma Dodd\inst{1} \and Amina Helmi\inst{1}
}

\institute{Kapteyn Astronomical Institute, University of Groningen,
          Landleven 12, 9747 AD Groningen, The Netherlands\\
          \email{s.s.lovdal@rug.nl}
    \and
          School of Natural Sciences, Institute for Advanced Study, 1 Einstein Drive, Princeton, NJ 08540, USA\\
}


\abstract
{Merger debris is expected to populate the stellar haloes of galaxies. In the case of the Milky Way, this debris should be apparent as clumps in a space defined by the orbital integrals of motion of the stars.}
{Our aim is to develop a data-driven and statistics-based method for finding these clumps in integrals-of-motion space for nearby halo stars and to evaluate their significance robustly.}
{We used data from \textit{Gaia} EDR3, extended with radial velocities from ground-based spectroscopic surveys, to construct a sample of halo stars within 2.5 kpc from the Sun. We applied a hierarchical clustering method that makes exhaustive use of the single linkage algorithm in three-dimensional space defined by the commonly used integrals of motion energy $E$, together with two components of the angular momentum, $L_z$ and $L_\perp$. To evaluate the statistical significance of the clusters, we compared the density within an ellipsoidal region centred on the cluster to that of random sets with similar global dynamical properties. By selecting the signal at the location of their maximum statistical significance in the hierarchical tree, we extracted a set of significant unique clusters. By describing these clusters with ellipsoids, we estimated the proximity of a star to the cluster centre using the  Mahalanobis distance. Additionally, we applied the HDBSCAN clustering algorithm in velocity space to each cluster to extract subgroups representing debris with different orbital phases.}
{Our procedure identifies 67 highly significant clusters ($ > 3\sigma$), containing $12$\% of the sources in our halo set, and $232$ subgroups or individual streams in velocity space. In total, 13.8\% of the stars in our data set can be confidently associated with a significant cluster based on their Mahalanobis distance. 
Inspection of the hierarchical tree describing our data set reveals a complex web of relations between the significant clusters, suggesting that they can be tentatively grouped into at least six main large structures, many of which can be associated with previously identified halo substructures, and a number of independent substructures. This preliminary conclusion is further explored in a companion paper, in which we also characterise the substructures in terms of their stellar populations.
}
{Our method allows us to systematically detect kinematic substructures in the Galactic stellar halo with a data-driven and interpretable algorithm. The list of the clusters and the associated star catalogue are provided in two tables in electronic format. 
}


\maketitle
%

\section{Introduction}
According to the $\Lambda$ cold dark matter model, galaxies grow hierarchically by merging with smaller structures \citep{springel2005simulations}. In the Milky Way, footprints from such events have been predicted to be observable particularly in the stellar halo \citep[see e.g.][]{Helmi2020streams} because mergers generally deposit their debris in this component. Wide-field photometric surveys have enabled detection of large spatially coherent overdensities and numerous stellar streams in the outer halo (here defined to be beyond 20 kpc from the Galactic centre); see for example \citet{Ivezic2000,Yanny2000,Majewski2003ApJ,Belokurov2006,Bernard2016} and also \citet{Mateu2018} for a compilation. In the inner halo - here defined to be a region of $\sim$20~kpc radial extent from the Galactic centre, roughly corresponding to the region probed by orbits of the stars crossing the solar vicinity - the advent of large samples with phase-space information from the {\it Gaia} mission \citep{prusti2016gaia,Brown2018} enabled the discovery of several kinematic substructures, particularly in the vicinity of the Sun. One of the most significant substructures both because of its extent and its importance is the debris from a large object accreted roughly 10 Gyr ago, named {\it Gaia}-Enceladus \citep[][see also \citealt{Belokurov2018}]{helmi2018merger}, and which has been estimated to comprise $\sim 40$\% of the halo near the Sun \citep{2020MNRAS.492.3631M,Helmi2020streams}. In addition, a hot thick disk \citep{helmi2018merger,Dimatteo2019A&A...632A...4D}, now also known as the ``splash'' \citep{Belokurov2020} is similarly dominant amongst stars near the Sun with halo-like kinematics. Another possible major building block of the inner Galaxy is the so-called Heracles-Kraken system, associated with a population of low-energy globular clusters and embedded in the inner parts of the Milky Way \citep[see][]{Massari2019A&A...630L...4M, 2019MNRAS.486.3134K, 2020MNRAS.498.2472K, 2020MNRAS.493.3363H, 2021MNRAS.500.1385H}. The remaining substructures are more modest in size and likely correspond to much smaller accreted systems \citep[see e.g.][]{Yuan2020ApJ...891...39Y}.

The main goal in the identification of these substructures is to determine and characterise the merger history of the Milky Way. The identification of the various events allows placing constraints on the number of accreted galaxies, their time of accretion, and their internal characteristics, such as their star formation and chemical evolution history, their mass and luminosity, and the presence of associated globular clusters \citep[e.g.][]{myeong2018discovery,Massari2019A&A...630L...4M}. The characterisation of the building blocks is interesting from the perspective of cosmology and galaxy formation because it allows the determination of the luminosity function across cosmic time, for example. Furthermore, the star formation histories and chemical abundance patterns in accreted objects permit distinguishing types of enrichment sites or channels (e.g. super- or hypernovae), as well as the initial mass function in different environments. Importantly, many of the high-redshift analogues of the accreted galaxies will not be directly observable in situ because of their intrinsically faint luminosity, even with the James Webb Space Telescope (JWST), and access to this population will therefore likely only remain available in the foreseeable future through studies of nearby ancient stars. The ambitious goals of
Galactic archaeology require that we are able to identify substructures in a statistically robust way to ultimately be able to assess incompleteness or biases, and to establish which stars are likely members of the various objects identified as this is necessary for their characterisation (eventually through detailed spectroscopic follow-up). 

Theoretical models and numerical simulations have shown that accreted objects will preserve their coherent orbital configuration long after the structure is completely phase mixed \citep{helmi2000mapping}, even in the fully hierarchical regime of the cosmological assembly process \citep{Gomez2013MNRAS.436.3602G,2019MNRAS.490L..32S}.
One of the best ways to trace accretion events is to observe clustering in the integrals of motion describing the orbits of the stars. In an axisymmetric time-independent potential, frequently used integrals of motion are the energy $E$ and angular momentum in the $z$-direction $L_z$. 
The component of the angular momentum $L_\perp$ (=$\sqrt{L_x^2+L_y^2}$) is often used as well because stars originating in the same merger event are expected to remain clustered in this quantity as well, even though it is not being fully conserved \citep[see e.g.][]{helmi1999debris}. Similarly, action space can be used \citep[see e.g.][]{myeong2018discovery}, with the advantage that actions are adiabatic invariants, and they are less dependent on distance selections \citep{2022MNRAS.510.5119L}. The actions have the drawback, however, that they are more difficult to determine for a generic Galactic potential \citep[except in approximate form, see e.g.][]{Binney2016MNRAS.456.1982B,2019MNRAS.482.1525V}.

Establishing the statistical significance of clumps identified by clustering algorithms in these subspaces is not trivial. 
Thus far, many works have used either manual selection \citep[e.g.][]{naidu2020evidence} or clustering algorithms in integrals-of-motion space successfully to identify stellar streams in the Milky Way halo \citep{koppelman2019multiple,  borsato2020identifying}. 
In recent years, works using more advanced machine learning to find clusters have also been published \citep{2018ApJ...863...26Y, 2018MNRAS.475.1537M, borsato2020identifying}. In the case of manual selection, mathematically establishing the validity is difficult and often bypassed. 
For machine learning and advanced clustering algorithms, the results are generally hard to interpret, especially in the case of unsupervised machine learning, where there are no ground-truth labels. Although training and testing is possible via the use of numerical simulations \citep[see][]{2020ApJS..246....6S,2020A&A...636A..75O}, these simulations have many limitations themselves, and there is no guarantee that the results obtained can be extrapolated to real data sets. 
Furthermore, the vast majority of clustering algorithms require a selection of parameters that have a deterministic impact on the results \citep[see e.g.][and references therein]{rodriguez2019clustering}. At the same time, it is very difficult to physically motivate the choice of some parameters for a given data set. 

We present a data-driven algorithm for clustering in integrals-of-motion space, relying on a minimal set of assumptions. We also derive the statistical significance of each of the structures we identified, and define a measure of closeness between each star in our data set and any of the substructures. Our paper is structured as follows: Section~\ref{sec:data} describes the construction of our data set and the quality cuts we impose, while Section~\ref{sec:methods} covers the technical details of the clustering algorithm. In Section~\ref{sec:results} we evaluate the statistical significance of the clusters we found and present some of their properties, including their structure in velocity space. For each star in our data set, we also provide a quantitative estimate that it belongs to any of the statistically significant clusters. The results are interpreted and discussed in Section~\ref{sec:discussion}. We also explore here possible relations between the various extracted clusters and make a comparison to previous work. The first conclusions of our study are summarised in Section~\ref{sec:conclusions}. 
Since our analysis is based on dynamical information alone, this does not fully reveal the origin of the identified clusters by itself \citep[e.g. accreted vs. in situ, see][or their relation, see \citealt{2020A&A...642L..18K}]{2017A&A...604A.106J}.
 We devote paper II \citep{2022arXiv220102405R} to the characterisation and nature of the structures identified in this work.

\section{Data}\label{sec:data}
The basis of this work is the {\it Gaia} EDR3 RVS sample, that is, stars whose line-of-sight velocities were measured with the radial velocity spectrometer (RVS) included in the {\it Gaia} Early Data Release 3 \citep[EDR3,][]{2021A&A...649A...1G}, supplemented with radial velocities from ground-based spectroscopic surveys. In particular, we extend the {\it Gaia} RVS sample with data from the sixth data release of the Large Sky Area Multi-Object Fiber Spectroscopic Telescope (LAMOST DR6) low-resolution \citep[][]{2020ApJS..251...27W} and medium-resolution \citep[][]{2019RAA....19...75L} surveys, the sixth data release of the RAdial Velocity Experiment (RAVE DR6) \citep[][]{2020AJ....160...82S, 2020AJ....160...83S}, the third data release of the Galactic Archaeology with HERMES (GALAH DR3) \citep[][]{2021MNRAS.506..150B}, and the 16th data realease of the Apache Point Observatory Galactic Evolution Experiment (APOGEE DR16) \citep[][]{2020ApJS..249....3A}. We performed spatial cross-matching between these survey catalogues using the {\tt TOPCAT/STILTS} \citep[][]{2005ASPC..347...29T, 2006ASPC..351..666T} {\tt tskymatch2} function. We allowed a matching radius between sky coordinates up to $5$~arcsec after transforming J2016.0 {\it Gaia} coordinates to J2000.0, although the average distance between matches is $\sim$0.15~arcsec, and $95\%$ are below $0.3$~arcsec. For the LAMOST low-resolution survey, we applied a $+7.9$~km/s offset to the measured velocities according to the LAMOST DR6 documentation release\footnote{http://dr6.lamost.org/v2/doc/release-note}. Following \citet[][]{koppelman2019characterization}, and based on the accuracy of their line-of-sight velocity measurements, we first considered GALAH, then APOGEE, then RAVE, and finally LAMOST while extending the RVS sample. 
We also imposed a maximum line-of-sight velocity uncertainty of $20$ km/s. This maximum radial velocity error introduces a small bias against very metal-poor stars in the LAMOST-LRS data set. The final extended catalogue contains line-of-sight velocities for 10,629,454 stars with \texttt{parallax\_over\_error}~$>5$.

We consider the local halo as stars located within $2.5$ kpc, where the distance was computed by inverting the parallaxes after correcting for a zero-point offset of $0.017$~mas. We require high-quality parallaxes according to the criterion above and low renormalised unit weight error (\texttt{ruwe} $<1.4$). We assumed $V_{\textrm{LSR}} = 232$ km/s, a distance of $8.2$ kpc between the Sun and the Galactic centre \citep{mcmillan2016mass}, and used ($U_\odot, V_\odot, W_\odot$) = ($11.1, 12.24, 7.25$) km/s for the peculiar motion of the Sun  \citep{schonrich2010local}. 

We identified halo stars by demanding $|{\bf V}-{\bf V}_{\textrm{LSR}}| > 210 $ km/s, similarly to \citet{koppelman2018one, koppelman2019characterization}. This cut is not too conservative and allows for some contamination from the thick disk. Additionally, we removed a small number of stars whose total energy computed using the Galactic gravitational potential described in the next section is positive. The resulting data set contains $N=51671$ sources, and we refer to this selection as the halo set.

The relative parallax uncertainty of the stars in this halo sample is lower than 20\%. For example, the median \texttt{parallax\_error/parallax} for the {\it Gaia} RVS sample is $\sim 2.4$\%, while for the full extended sample ({\it Gaia} RVS plus ground-based spectroscopic surveys), it is $\sim 3.7$\%. The radial velocities of the stars from the {\it Gaia} RVS sample have much lower uncertainties than the cut of 
20~km/s, with a median line-of-sight velocity uncertainty of $1$~km/s. The full extended (halo) sample has a median line-of-sight velocity uncertainty of $6.7$~km/s, driven mostly by the higher uncertainties associated with the LAMOST DR6 low-resolution radial velocities. 
These uncertainties (as well as those in the proper motions) drive the uncertainties in the total velocities of the stars in our halo sample, as well as in their integrals of motion (see Sec.~\ref{subsec:iom_clustering}). The median uncertainty in the total velocity, $v$, is 4.1~km/s for stars from the {\it Gaia} RVS sample and 9.1~km/s for those from the extended sample.

\begin{figure*}[t!]
    \resizebox{\hsize}{!}
        {\includegraphics[width=\textwidth]{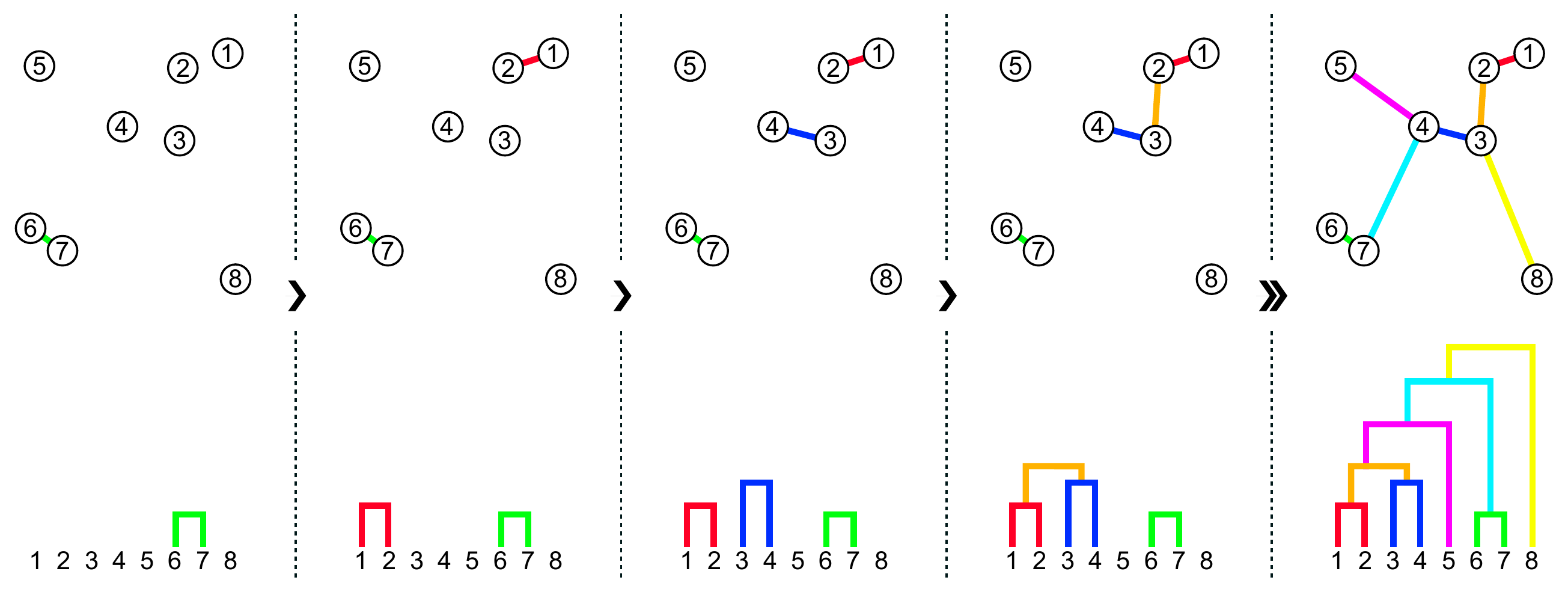}}
        \caption{Single linkage algorithm applied on a two-dimensional example. Each step of the algorithm forms a new group by connecting the two closest data points not yet in the same cluster, where each data point is considered a singleton group initially. The resulting merging hierarchy is visualised as a dendrogram at the bottom of each panel.}
    \label{fig:single_linkage_example}
\end{figure*} 

\section{Methods}\label{sec:methods}

One of the primary goals of this work is to develop a data-driven algorithm, where the extracted structures are statistically based and easily interpretable. At the same time, we desire a method that detects more than the most obvious clusters, but that is ideally able to scan through the data set without missing any signal.
 
In what follows, we use the following notation: ${\bf x}$ for a data point, or star, in our clustering space. The dimensionality of our data space is denoted by $n$, $N$ is the number of stars in the data set after applying quality cuts, and as described in Sec.~\ref{sec:data}, we call this selection the halo set. $C_i$ denotes a connected component in the halo set, having been formed at step $i$ of the clustering algorithm. Each $C_i$ is a candidate cluster for which we wish to evaluate the statistical significance. We denote the number of members of a candidate cluster $N_{C_i}$. 

We consider the data points as such, without taking measurement uncertainties into account. While this would technically be possible, the quality cuts we impose on the data are strict enough to result in reasonably robust outcomes, and we leave it to future work to extend the method by taking uncertainties into account in a probabilistic way.

\subsection{Clustering algorithm: Single-linkage}

A range of clustering algorithms is available, but we desire maximum control over the process, in combination with an exhaustive extraction of information in the data set. This is especially important as we are dealing with unsupervised machine learning, therefore there is no ground truth to verify our findings with from a computational point of view.

Our clustering algorithm will have to deal with a potentially large amount of noise in search for significant clusters, which rules out clustering algorithms that assume that all data points belong to a cluster. Some options that can handle noise, and which have also been used to extract clusters in integrals-of-motion space, would be the friends-of-friends algorithm \citep{efstathiou1988gravitational}, used for instance in \citet{helmi2000mapping}, {\tt DBSCAN} \citep{ester1996density}, used by \citet{borsato2020identifying}, and {\tt HDBSCAN} \citep{campello2013density}, as in \citet{koppelman2019multiple}. The main problem with the first two is that they require specifying some static parameters, for example a distance threshold for data points of the same cluster. As there is a gradient in density in our halo set (especially, fewer stars with lower binding energies), using static parameters for this will only work well on some parts of the data space, unless we first apply some advanced non-linear transformations. {\tt HDBSCAN} is able to extract variable density clusters, but in addition to having a slight black-box tendency in the application, the output is also dependent on some user-specified parameters. We wish to make as few assumptions as possible about the properties of the clusters and also desire full control over the clustering process, therefore we consider a simpler option.

The single-linkage algorithm is a hierarchical clustering method that only requires a selection of distance metric \citep{everitt2011cluster}. We used standard Euclidean distance to this end. At each step of the algorithm, it connects the two groups with the smallest distance between each other, defined as the smallest distance between two data points not yet in the same group, where each data point is considered a singleton group initially. Each merge, or step $i$ of the algorithm, corresponds to a new connected component in the data set. This is illustrated in Fig.~\ref{fig:single_linkage_example}. Here single-linkage is applied on a two-dimensional example, where the top part of each panel illustrates the merging process of the algorithm, and in the bottom we illustrate the resulting merging hierarchy in a dendrogram. If the data set is of size $N,$ the algorithm performs $N-1$ steps in total as it continues until the full data set has been linked. Hence, the algorithm is also closely related to graph theory, as the result after the last merge is equivalent to the minimum spanning tree \citep{gower1969minimum}. From a computational point of view, the series of connected components that is obtained also corresponds to the set of every potential cluster in the data, under the assumption that the most likely clusters are the groups of data points with the smallest distance between each other, without assuming any specific cluster shape.

\begin{figure*}[t!]
    \resizebox{\hsize}{!}
        {\includegraphics[width=\textwidth]{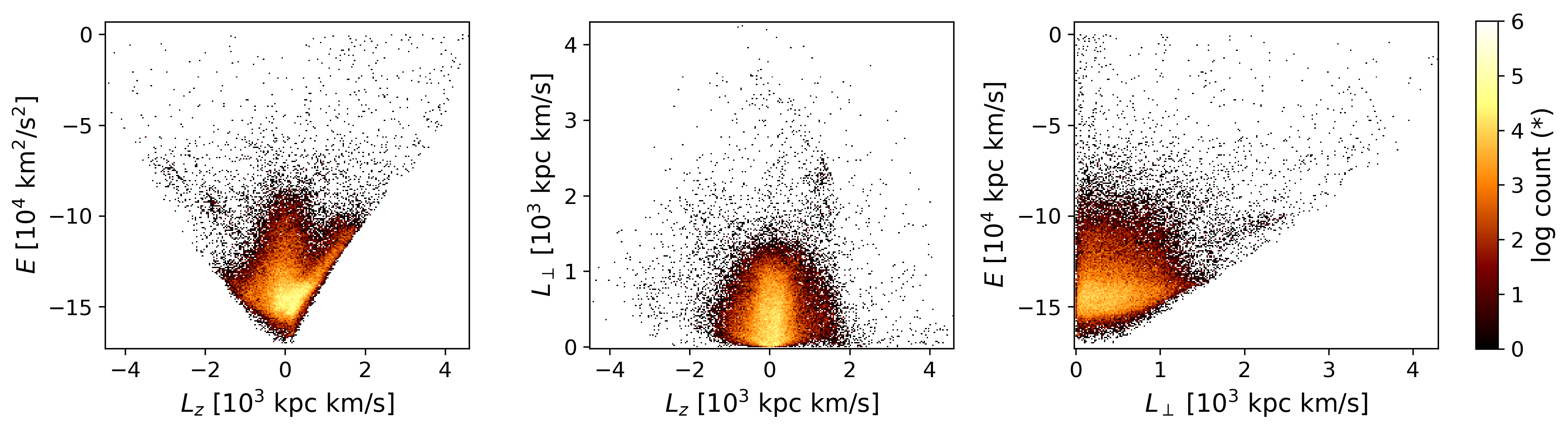}}
        \caption{Distribution of stars in our halo sample in integrals-of-motion space. The single-linkage algorithm identifies clusters in the three-dimensional space resulting from the combination of our three clustering features, namely energy $E$, and two components of the angular momentum $L_z$ and $L_\perp$.}
    \label{fig:halo_set_subspaces}
\end{figure*}

The core idea of our clustering method is to apply the single-linkage algorithm to the halo set and subsequently evaluate each connected component (or candidate cluster $C_i$, formed at step $i$ of the algorithm) by a cluster quality criterion. The clusters that exhibit statistical significance according to the selected criterion are accepted. In this way, the method is also able to handle noise because the data points that do not belong to any cluster displaying a high statistical significance are discarded.

\subsection{Clustering in integrals-of-motion space}
\label{subsec:iom_clustering}

As described earlier, to identify merger debris, we relied on the expectation that it should be clustered in integrals-of-motion space. As integrals of motion, we used three typical quantities: The angular momentum in $z$-direction $L_z$, the perpendicular component of the total angular momentum vector $L_{\perp}$, and total energy $E$. While $L_z$ is truly conserved in an axisymmetric potential, $L_{\perp}$ typically varies slowly, retaining a certain amount of clustering for stars on similar orbits as those originating in the same accretion event, although not being fully conserved. The total energy $E$ is computed as
\begin{equation}
    E = \frac{1}{2} v^2 + \Phi({\bf r})
,\end{equation}
where $\Phi({\bf r})$ is the Galactic gravitational potential at the location of the star. For this we used the same potential as in \citet{koppelman2019characterization}: A Miyamoto-Nagai disk, Hernquist bulge, and Navarro-Frank-White halo with parameters $(a_d, b_d) = (6.5, 0.26)$ kpc, $M_{d}=9.3\times 10^{10} M_\odot$ for the disk, $c_b = 0.7$ kpc, $M_{b}=3.0 \times 10^{10} M_\odot$ for the bulge, and $r_s=21.5$ kpc, $c_h$=12, and $M_{\rm halo}=10^{12} M_\odot$ for the halo. While the choice of potential influences the absolute values we computed for total energy, the difference in the distributions of data points between two reasonably realistic potentials is negligible in the clustering space, as shown in Section \ref{sec:discussion}.

We scaled each of the integrals of motion, or `features', linearly to the range $[-1, 1]$ using a set reference range of $E=[-170000, 0]$ km$^2$/s$^2$, $L_\perp=[0, 4300]$ kpc km/s, and  $L_z=[-4500,4600]$ kpc km/s, roughly corresponding to the minimum and maximum values in the halo set.
Our equal-range scaling also implies that each of the three features is considered equally important in a distance-based clustering algorithm. The typical (median) errors in these quantities are much smaller than the range they cover. In the full halo sample, for example, for the energy $\langle\sigma_E\rangle \sim 1052$ km$^2$/s$^2$; and for the angular momenta  $\langle\sigma_{L_z}\rangle \sim 61$~kpc~km/s and $\langle\sigma_{L_\perp}\rangle \sim 46$ kpc~km/s, , while for the subset with Gaia RVS velocities, the typical uncertainties are a factor 2.5 -- 3.5 smaller.
The halo set visualised for all combinations of the clustering features is shown in Fig.~\ref{fig:halo_set_subspaces}.

\begin{figure*}
    \resizebox{\hsize}{!}
        {\includegraphics[width=\textwidth]{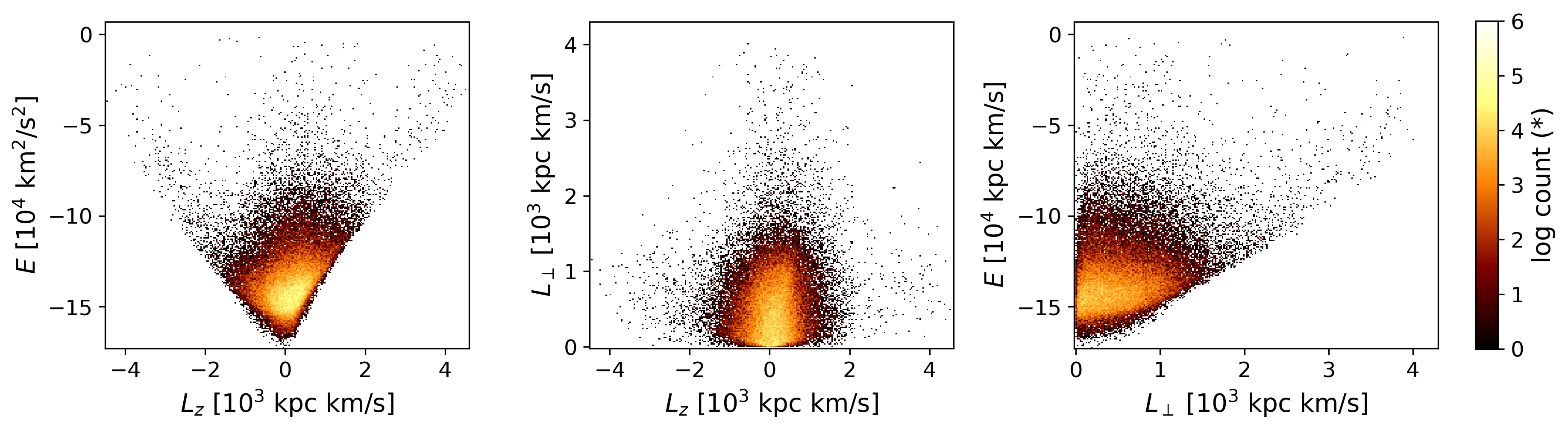}}
        \caption{Example of the distribution of points in integrals of motion for one of the random halo sets (number 1 out of $N_{\rm art}=100$) obtained by re-shuffling the  velocity components. See for comparison Fig.~\ref{fig:halo_set_subspaces}.}
        \label{fig:art_data}
\end{figure*}

\subsection{Random data sets}

To assess the statistical significance of the outcome of our clustering algorithm, we used random data sets. To this end, we created a reference halo set using our existing data set, but recomputed the integrals of motion using random permutations of the $v_y$ and $v_z$ components, similarly to \citet{helmi2017box}. This created an artificial data set with similar properties to the observed data, but where the correlations in the velocity components and hence structure in integrals of motion space is broken up. Specifically, we scrambled the velocity components of all stars with  $|{\bf V}-{\bf V}_{\textrm{LSR}}| > 180 $ km/s (a slightly more relaxed definition of the halo, comprising $75149$ sources of our original data set). The majority of these stars with scrambled velocities satisfied the criterion we used to kinematically define halo stars, $|{\bf V}-{\bf V}_{\textrm{LSR}}| > 210 $ km/s. For each artificial halo, we therefore sampled $N$ stars satisfying the above-mentioned criterion and recomputed the clustering features, where $N$ is the number of stars in our original halo set. We normalised these features with the same scaling as for the original halo set, such that the mapping from absolute to scaled values was identical for the real and artificial data.

We generated $N_{\rm art}=100$ realisations of this artificial halo as reference. An example artificial data set is displayed in Fig.~\ref{fig:art_data}, where an arbitrarily chosen realisation is visualised. By comparing to Fig.~\ref{fig:halo_set_subspaces}, we see that the two data sets have very similar characteristics, but that the substructure visible by eye in the original halo set has been diluted.

\section{Results} \label{sec:results}

\subsection{Identification of clusters}

We applied the algorithm described above on our halo set. We only investigated candidate clusters with at least ten members, evaluating them according to the procedure described in the next section, as this is the smallest group that would be statistically significant assuming Poissonian statistics together with the significance level we adopted.

\subsubsection{Evaluating statistical significance}
\label{subsec:significance}

In order to examine the quality of each candidate cluster obtained by the linking process, we need a cluster evaluation metric. A challenge is that our three-dimensional clustering space has a higher density of stars with low energy and angular momentum than regions with higher energy, as shown in  Fig.~\ref{fig:halo_set_subspaces}. 
Therefore we used our randomised data sets to compare the observed and expected density for different regions. 
We thus examined the candidate clusters resulting from applying the single-linkage algorithm and assessed their statistical significance by computing the expected density of stars in a region, and comparing the difference between the observed and expected count in relation to the statistical error on these quantities.

\begin{figure*}[!t]
    \resizebox{\hsize}{!}
        {\includegraphics[trim=0 0.2cm 0 0, clip,width=0.8\textwidth]{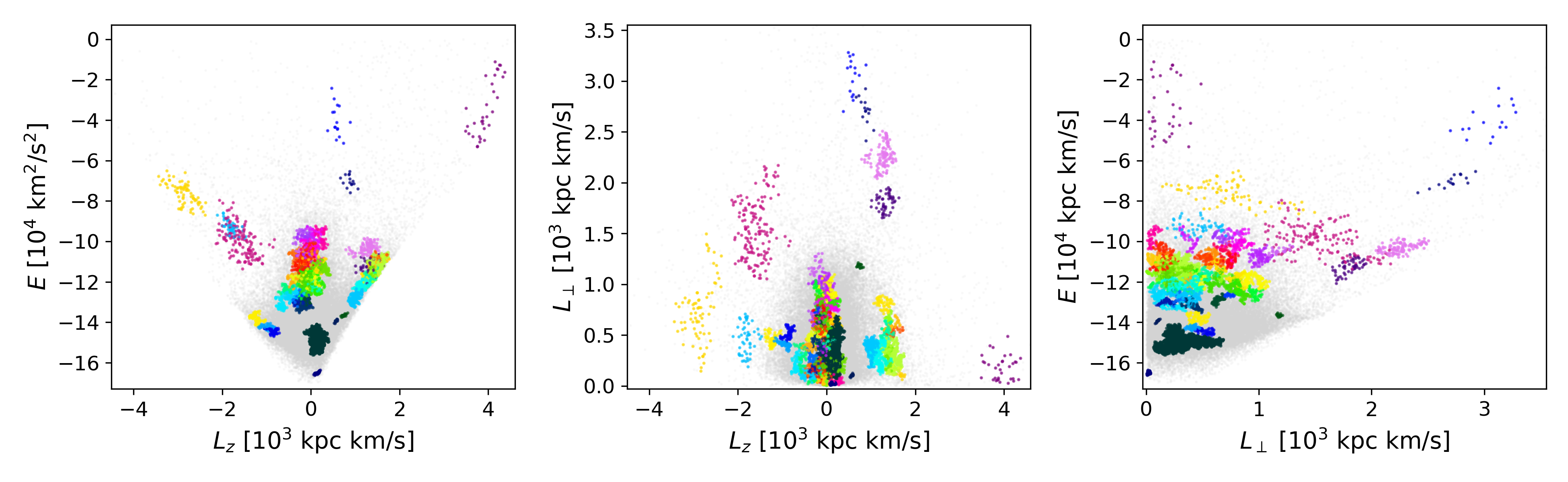}}
        {\includegraphics[trim=0 0 0 0.4cm, clip, width=\textwidth]{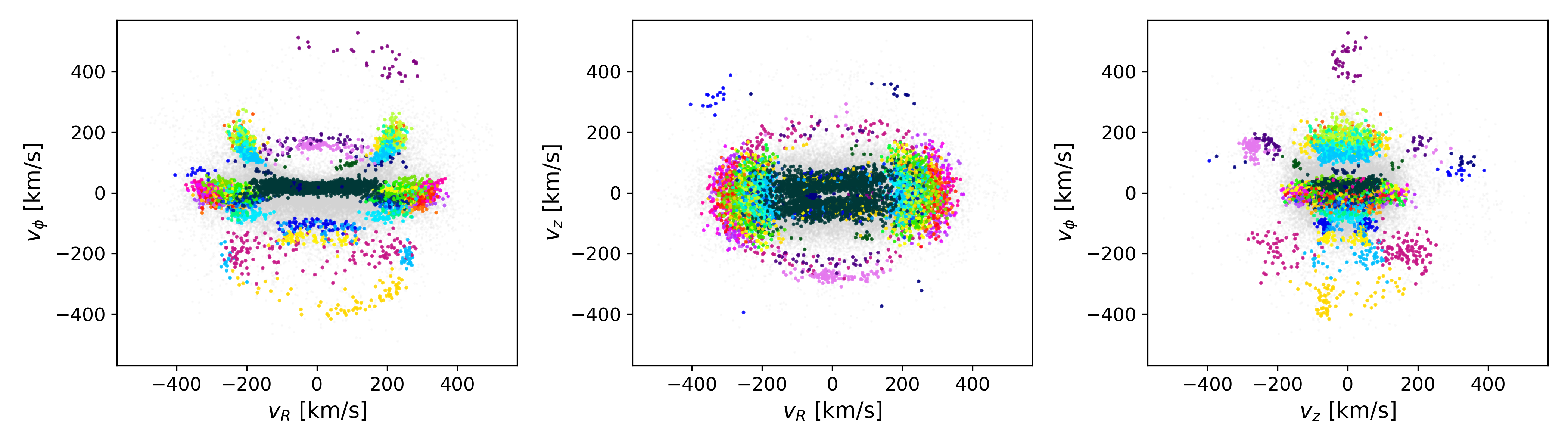}}
        \caption{Clusters extracted by the algorithm, visualised in integrals of motion (top row) and in velocity space (bottom row). The different colours indicate the stars associated with the 67 different clusters we identified.
        }
    \label{fig:clusters_subspaces}
\end{figure*}

In order to compare the number of members of a candidate cluster $C_i$ to the expected count (obtained from our randomised sets), we determined the region in which $C_i$ resides.
To this end, we defined an ellipsoidal boundary around $C_i$ by applying principal component analysis (PCA) on the members. The standard equation of an $n$-dimensional ellipsoid centred around the origin is 
\begin{equation}
 \sum_{i=1}^{n} \frac{x_i^2}{a_i^2} = 1
,\end{equation}
where $a_i$ denotes the length of each axis.
The variance along each principal component of the cluster is given by the eigenvalues $\lambda_i$ of the covariance matrix, and we can define the length of each axis $a_i$ of the ellipsoid in terms of the number of standard deviations of spread along the corresponding axis. We consulted the $\chi^2$ distribution with three degrees of freedom (corresponding to our three-dimensional clustering space) and observed that $95.4\%$ of a three-dimensional Gaussian distribution falls within $2.83$ standard deviations of extent along each axis. This is the fraction of a distribution that directly corresponds to the length of two standard deviation axes in a univariate space. Hence, we chose the axis lengths to be $a_i = 2.83\sqrt{\lambda_i}$. 
The choice to cover $95.4\%$ of the distribution provides a snug boundary around the data points that is neither too strict nor includes too much empty space.

We then computed the number of stars falling within the ellipsoidal cluster boundary by analysing the PCA transformation of $C_i$ and mapping the stars of the data sets to the PCA space defined by $C_i$ by subtracting the means and multiplying each data point by the eigenvectors of the covariance matrix. Hereby we obtained a mean centred and rotated version of the data in which the direction of maximum variance aligns with the axis of the coordinate system.

We computed both the average number of stars from the artificial halo $\langle N_{C_i}^{\rm art}\rangle$ and the number of stars from the real halo set $N_{C_i}$ that fall within this boundary. The statistical significance of a cluster was then obtained by the difference between the observed and expected count, divided by the statistical error on both quantities. We required a minimum significance level of $3 \sigma$, defined by
\begin{equation}\label{eq:significant_req}
N_{C_i} - \langle N_{C_i}^{\rm art} \rangle > 3 \sigma_i,
\end{equation}
and where 
$\sigma_i = \sqrt{N_{C_i}+(\sigma^{\rm art}_{C_i})^2}$.
Here $\sigma^{\rm art}_{C_i}$ is the standard deviation in the number counts across our $100$ artificial halos.
We treated the observed data as having Poissonian properties, giving the statistical error on the observed cluster count as $\sqrt{N_{C_i}}$.

\subsubsection{Statistically significant groups}\label{subsec:signific}

Evaluating the statistical significance for all candidate clusters returned by the single linkage algorithm returned a set of clusters with a $3 \sigma$ significance at least, some of which were hierarchically overlapping subsets of each other. A structure is likely to display statistical significance starting from the core of the cluster, growing out to its full extent via a series of merges in the algorithm, each being deemed statistically significant. Similarly, a cluster with a dense core together with some neighbouring noise still displays significance if the core is dense enough. 

Under the hypothesis that the statistical significance increases while more stars of the same structure are being merged into the cluster and that the significance decreases when noise in the outskirts is added, final (exclusive) cluster labels were found by traversing the merging tree of the overlapping significant clusters and selecting the location at which they reached their maximum significance. In practice, this was done by iterating over the clusters ordered by descending significance, traversing its parent clusters upwards in the tree, and selecting the location where the maximum significance was reached. Nodes higher up in the merging hierarchy than the maximum significance for this path were removed because we did not aim for a path with a smaller maximum significance to override the selection by deeming a larger cluster on the same path to be a final cluster.

After finding a cluster in the tree at its maximum significance, we assigned the same label to all stars that belonged to this connected component according to the single linkage algorithm. 
Conveniently, if we were to chose a lower significance level than $3\sigma$, the change would simply extract additional clusters of a lower significance that do not overlap with those of higher significance.

\begin{figure}
\centering
\includegraphics[trim=0.3cm 0 1.5cm 0.2cm, clip,width=\hsize]{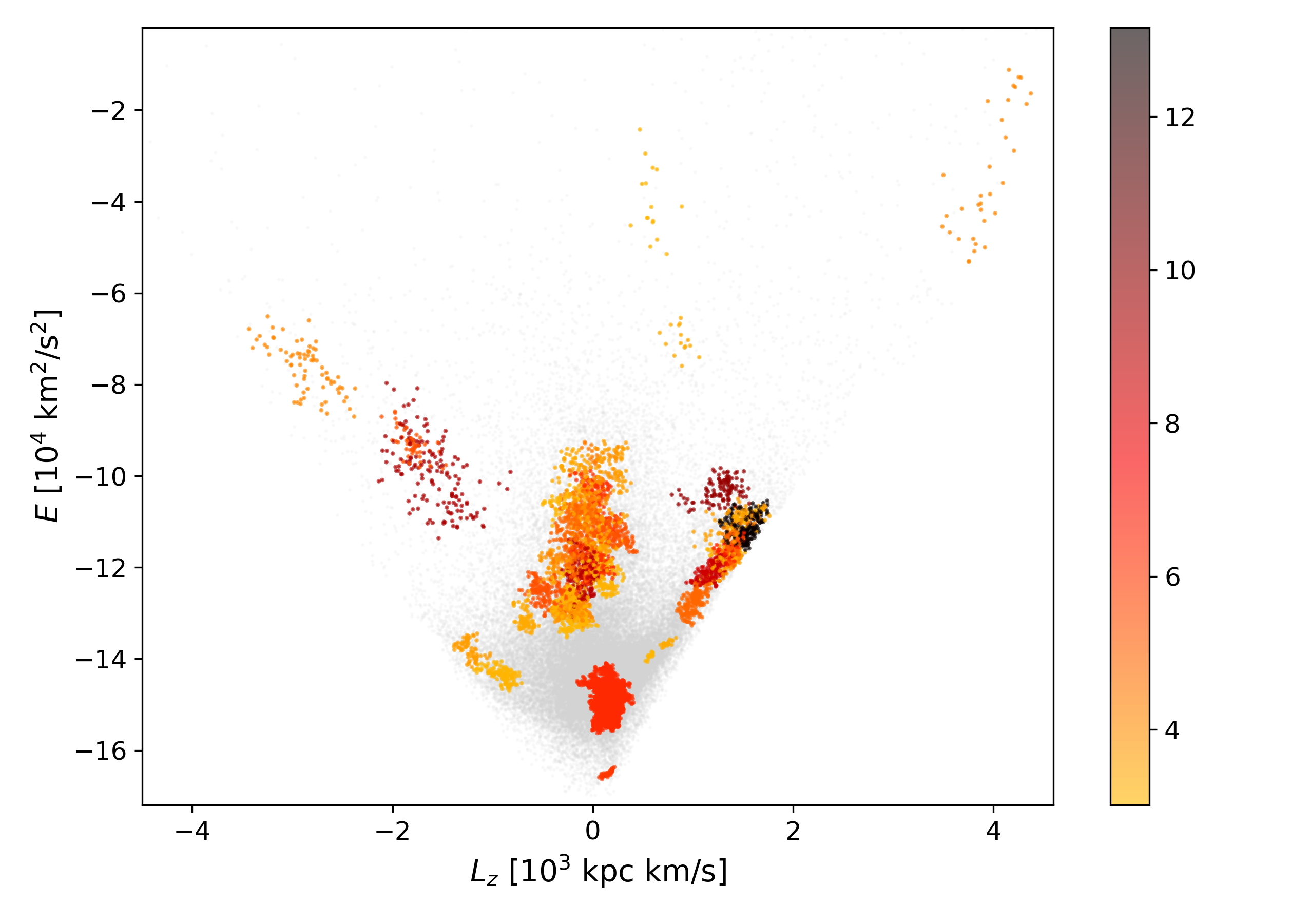}
  \caption{Distribution of the 67 high-significance clusters in the $E-L_z$ plane, colour-coded according to their statistical significance as computed by Eq.~\ref{eq:significant_req}.}
  \label{fig:significance_map}
\end{figure}

The results of the methodology described above are shown in Fig.~\ref{fig:clusters_subspaces}, which plots the distribution of the individual clusters identified for the various subspaces (top panel) and in velocity space (bottom panel). The colours here are for illustration purposes only, and there are enough clusters such that some of them are plotted in very similar shades.

One of clusters identified by the single-linkage algorithm (assigned label, or cluster ID $1$) is the globular cluster M4. This cluster has the lowest energy and is shown in dark blue in the top left panel of  Fig.~\ref{fig:clusters_subspaces}. Although all of its associated stars have nominally good parallax estimates, a subset has rather high correlation values in the covariance matrix provided by the {\it Gaia} database, particularly for terms involving the parallax (explaining finger-of-god features in velocity space). This finding prompted us to inspect all of the significant clusters, and we realised that cluster $67$ also has members with similarly high values of correlation
terms involving parallax (namely those with declination and $\mu_\delta$, in particular, most of its stars either have \texttt{dec\_parallax\_corr} that is below -0.15 or \texttt{parallax\_pmdec\_corr} above 0.15). These stars are also located towards the Galactic centre and anticentre, that is, they also lie in relatively crowded regions. 
The average value of the stars in cluster $67$  in \texttt{dec\_parallax\_corr} and \texttt{parallax\_pmdec\_corr} is significantly lower than that of any other cluster, resulting in an average \texttt{parallax\_err}=0.1, compared to an average value lower than 0.045 for all other clusters. To avoid confusion, we did not plot this cluster in Figure~\ref{fig:clusters_subspaces} and did not consider it further in our analysis.

Figure~\ref{fig:significance_map} shows the clusters colour-coded by their statistical significance. The algorithm identified $67$ clusters to which a total of $6209$ stars were assigned. The range of significance values is $[3.0, 13.2]$.
The figure shows that the clusters with the highest significance are located in regions of already known substructures, such as the Helmi Streams near $L_z \sim 1500$~kpc~km/s and $E \sim -10^5$ km$^2$/s$^2$ \citep{helmi1999debris,  koppelman2019characterization}.

\begin{figure}
\centering
\includegraphics[width=\hsize]{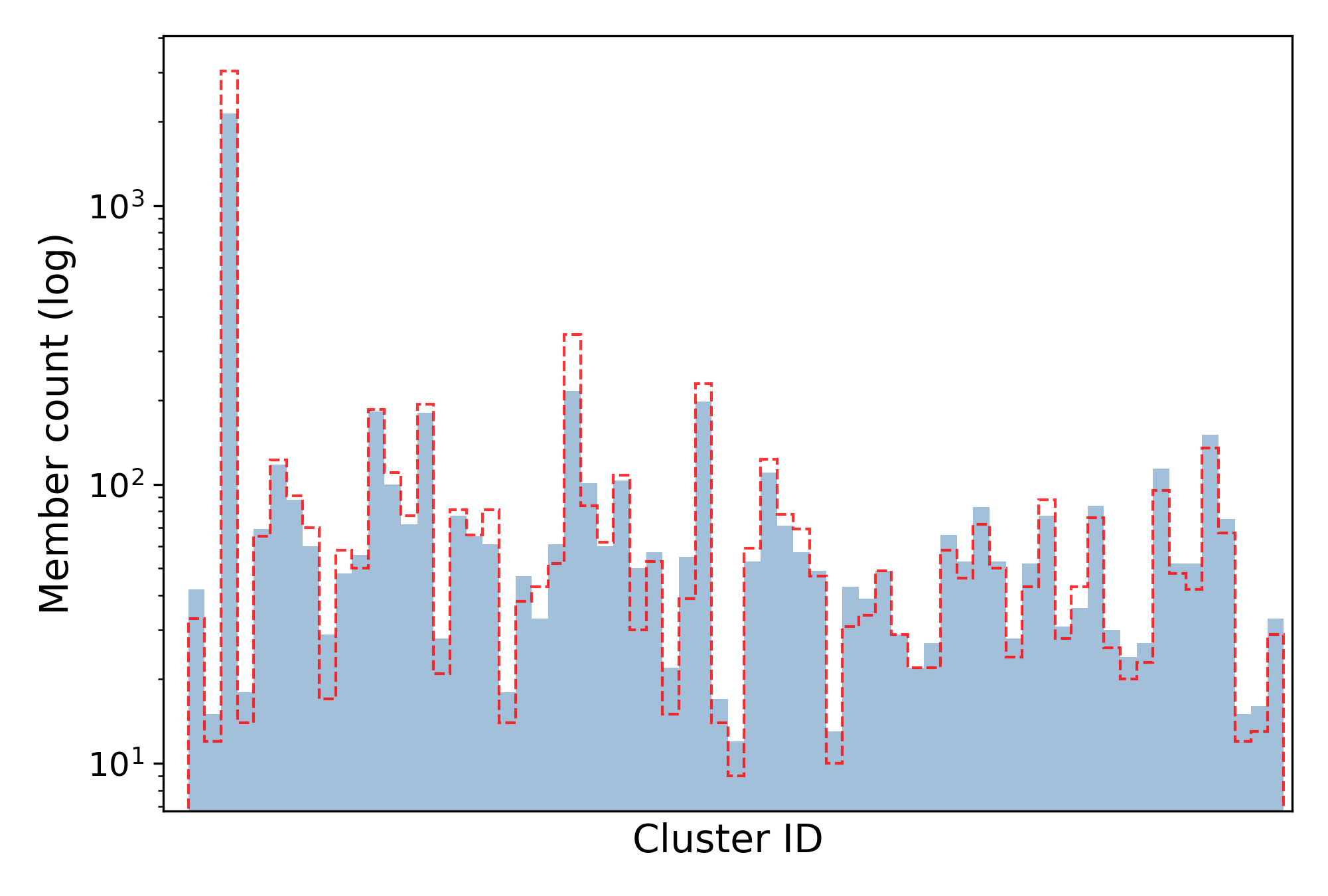}
  \caption{Histogram of the number of members per cluster. Members identified by the single-linkage algorithm are plotted in solid blue. The dashed red line indicates the sizes when original and additional plausible members are considered within a Mahalanobis distance of $D_{\rm cut} \sim 2.13$ from each cluster centre.}
  \label{fig:n_members_hist}
\end{figure}

In Fig.~\ref{fig:n_members_hist} we show the distribution of the number of members of the clusters identified by our procedure, plotted as the histogram with blue bars. The median number of stars in a cluster is $53$, and $2137$ stars are associated with the largest cluster. These numbers reflect the members identified by the linking process, which belong to some significant cluster $C_i$. For most clusters, it is possible to identify additional plausible members following the procedure described in Sec.~\ref{subsec:membership_probability}. The resulting cluster sizes are shown as the dashed red histogram in Fig.~\ref{fig:n_members_hist}.

\subsection{Extracting subgroups in velocity space}\label{subsec:clustering_vspace}

The velocity distributions of the extracted clusters contain more clues about their validity and properties because an accreted structure observed within a local volume is expected to display (sub)clumping in velocity space. This represents debris streams with different orbital phases.
 
We extracted these subgroups in velocity space by applying another round of clustering on each of the $67$ clusters. As the data set representing such a cluster in velocity space is far smaller and less complex than our original halo set, and because the subgroups in velocity space are often clearly separated clumps, we simply applied the HDBSCAN algorithm \citep{campello2013density} once for each cluster. This is a robust clustering algorithm that can extract variable density clusters and labels possible noise, while being able to handle various cluster shapes.

\begin{figure*}[]
    \centering
    \resizebox{0.90\hsize}{!}
        {\includegraphics[width=1.0\textwidth]{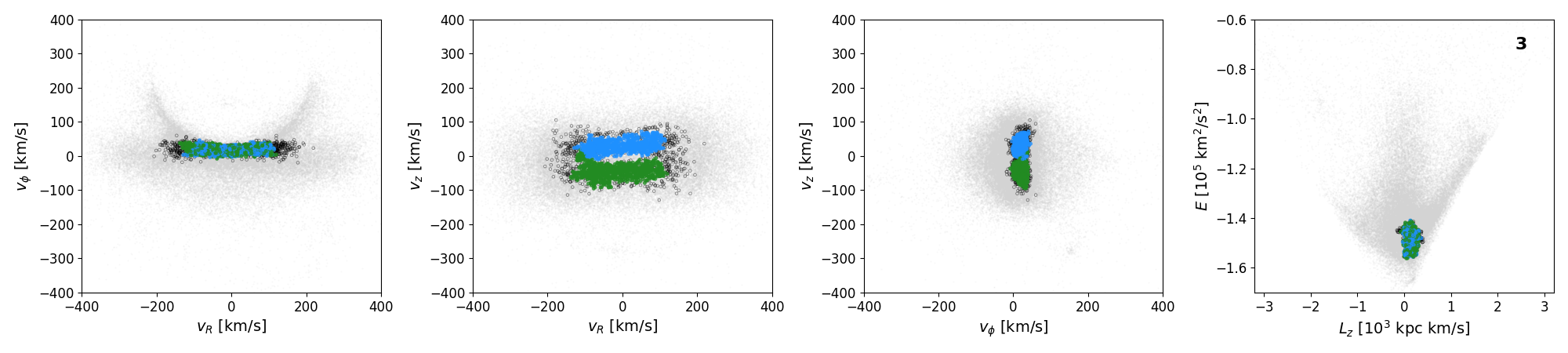}}
        {\includegraphics[width=0.9\textwidth]{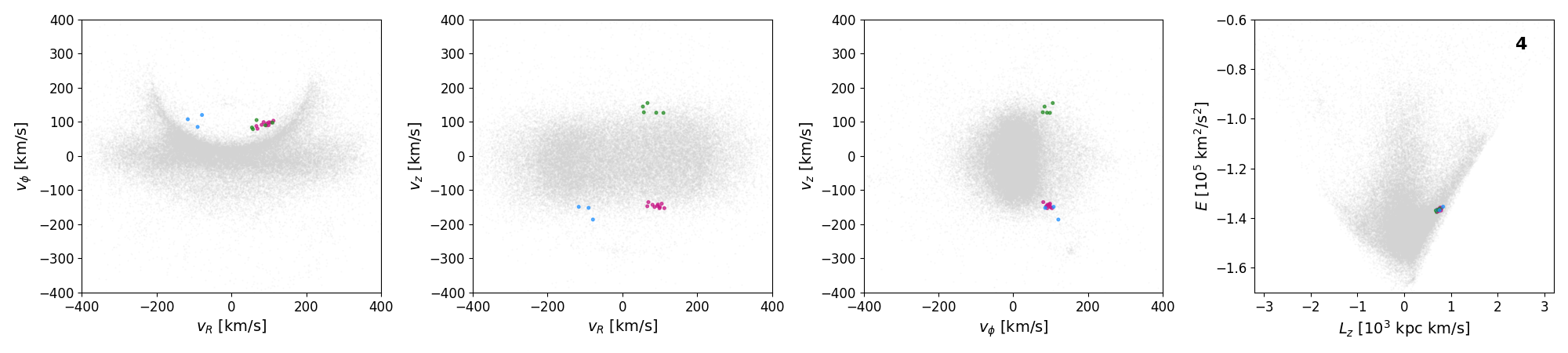}}
        {\includegraphics[width=0.90\textwidth]{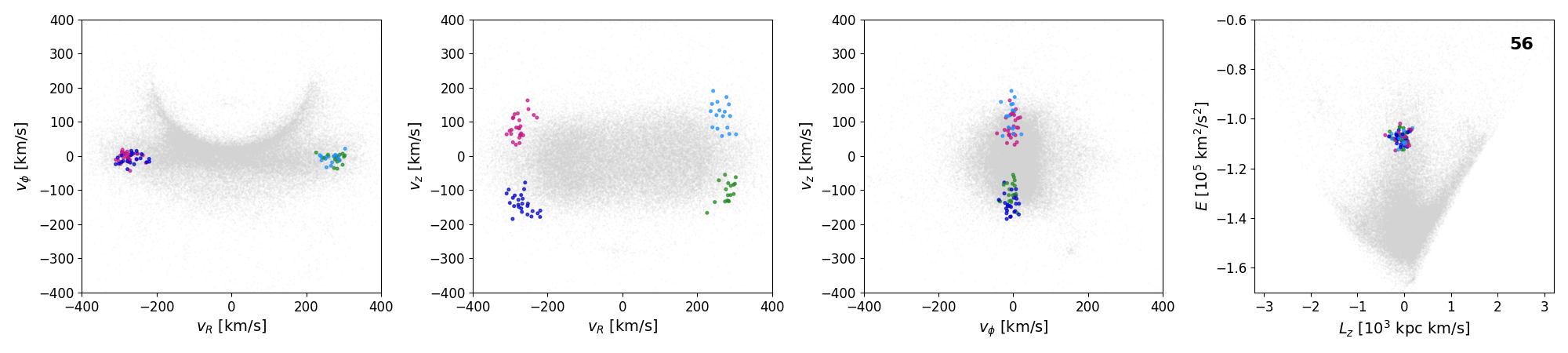}}
        {\includegraphics[width=0.90\textwidth]{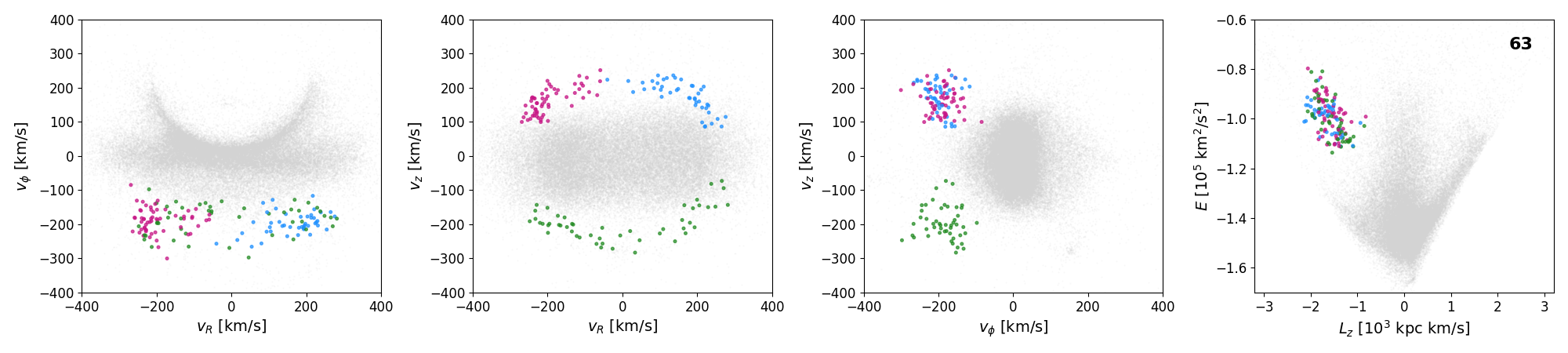}}
        {\includegraphics[width=0.90\textwidth]{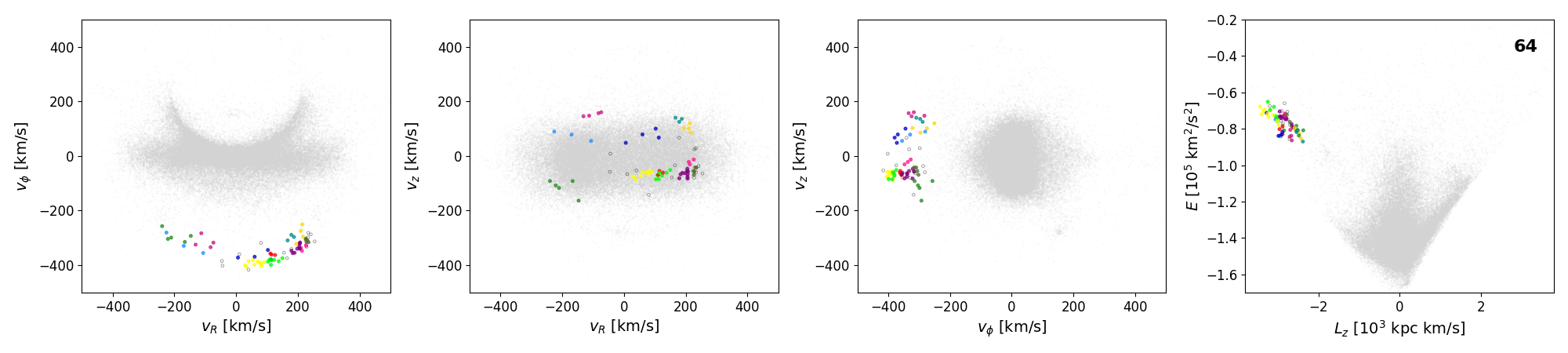}}
        \caption{Subgroups identified by HDBSCAN in velocity space for a selection of five significant clusters from the single-linkage algorithm, where noise as labelled by the algorithm is marked with open black circles. In each row, the three first panels display the projection of the subgroups in velocity space, and the fourth panel shows the location of the `parent' cluster in $E-L_z$ space; its ID is displayed in the upper right corner.}
    \label{fig:vspace_subgroups}
\end{figure*}
\begin{figure}
\centering
\includegraphics[trim=0.5cm 0 0.2cm 0 0, clip, width=\hsize]{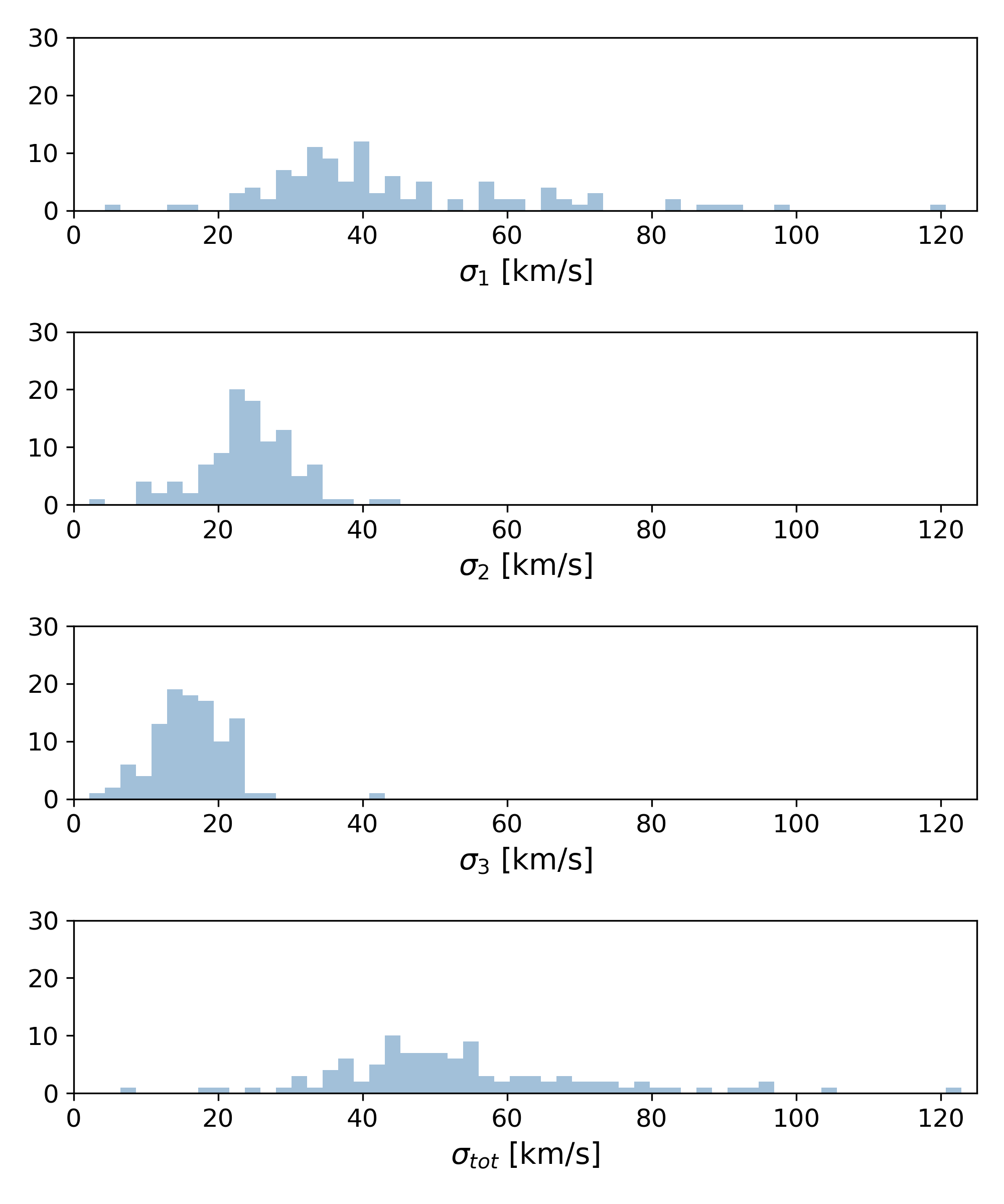}
  \caption{Velocity dispersions of each subgroup with at least $\text{ten}$ members identified by HDBSCAN, computed along the principal axes of the Cartesian velocity components (top three panels), and the total three-dimensional value (bottom panel).}
   \label{fig:v_space_dispersions}
\end{figure}

We applied HBDSCAN directly to the $v_R$, $v_\phi$ , and $v_z$ components of each cluster without scaling because the range for these values is already of the same magnitude. We set the parameter \texttt{min\_cluster\_size} (smallest group of data points that the algorithm accepts as an entity) to $5\%$ of the cluster size, with a lower limit of $\text{three}$ stars. We assigned \texttt{min\_samples} $=1$, which regulates how likely the algorithm is to classify an outlier as noise, with a lower value being less strict. The default mode of HDBSCAN does not allow for a single cluster to be returned, as its excess of mass algorithm may bias towards the root node of the data hierarchy. Because a single-linkage cluster might correspond to a single group in velocity space but this is not always the case, we circumvented this issue by setting the parameter \texttt{allow\_single\_cluster} to \texttt{True} only if the standard deviation in $v_R$, $v_\phi$ , and $v_z$ was small (below $30$ km/s) for at least two out of the three velocity directions. The idea is that if a cluster seems to be a single subgroup, it might display an elongated dispersion along one of these components at most, but if the dispersion is large for two or more out of the three, the cluster is likely to contain at least two kinematic subgroups.

In this way, the HDBSCAN algorithm extracted $232$ subgroups, where the number of subgroups per cluster varied in the range $[1, 12]$, with the mean and median being $3.5$ and $3,$ respectively. This is likely a lower limit to the true number of subgroups in velocity space, particularly because of the limitation in the number of stars, which prevented us from detecting streams with fewer than $\text{three}$ stars in the data set, but also due to the difficulty with defining optimal parameters for clustering in this space.

A series of examples of the output of HDBSCAN is displayed in Fig.~\ref{fig:vspace_subgroups}. Here each row reflects one cluster; the first three panels display the projection of the subgroups in velocity space, and the rightmost panel displays the location of the cluster in $E-L_z$ space. The cluster ID is indicated in the top right corner. Non-members are plotted in grey, and noise as labelled by HDBSCAN is shown with open black circles. 
The subgroups in velocity space for the statistically significant clusters are often quite distinct, for example in the case of clusters $3, 4,$ and $56$. Cluster $3$ is an example of a cluster that would be classified as a single subgroup if the parameter \texttt{allow\_single\_cluster} were statically \texttt{True}, demonstrating the necessity of our approach. Cluster $63$ is split into three subgroups, even though it could possibly be divided into either two or four groups as well. Cluster $64$ has the largest number of subgroups; it is divided into 12 small portions. 

The dispersion in the velocity components of these subgroups can be used to characterise them further. This was computed by applying PCA on the $v_x$, $v_y$ , and $v_z$ components of each subgroup with at least $\text{ten}$ members ($107$ in total), and measuring the standard deviation along the resulting principal components. A histogram of these dispersions is displayed in Fig.~\ref{fig:v_space_dispersions}. $\text{In 11}$ subgroups, the dispersion in the third principal component is smaller than $10$ km/s, and in 3 subgroups, it is below $5$ km/s. The lowest value of $2.5$ km/s is associated with the globular cluster M4 (cluster $1$) and the second smallest ($4.4$ km/s) with cluster $64$, which is shown as the yellow subgroup in the corresponding panel in Figure~\ref{fig:vspace_subgroups}.

The distribution of the three-dimensional velocity dispersion $\sigma_{tot}$ of the subgroups is shown in the bottom panel in Fig.~\ref{fig:v_space_dispersions}, where the lowest value is again obtained for the globular cluster M4. A single subgroup, the green subgroup of cluster $63$ in Fig.~\ref{fig:vspace_subgroups}, is truncated from Fig.~\ref{fig:v_space_dispersions} as it has a total velocity dispersion of $186$ km/s.

\subsection{Evaluating the proximity of stars to clusters}\label{subsec:membership_probability}

Now that we have obtained our cluster catalogue, we would like to obtain estimates for the likelihood of any individual star to belong to some specific cluster. This was done both in order to identify possible new members of a cluster and to find the most likely members for observational follow-up, for instance.

To do this, we described each cluster in integrals-of-motion space as a Gaussian probability density, defined by the mean and covariance matrix of the associated stars identified by the single-linkage algorithm. The idea is that the core of the overdensity is located around the mean of the Gaussian, and less likely members are located in the outskirts of the distribution. We then computed the location of a data point ${\bf x}$ within the Gaussian distribution by calculating its Mahalanobis distance $D,$
\begin{equation}
    D = \sqrt{({\bf x}-{\boldsymbol \mu})^T \Sigma ^{-1} ({\bf x}-{\boldsymbol \mu})}\label{eq:mah_distance}
,\end{equation}
where ${\boldsymbol  \mu}$ is the mean of the cluster distribution and $\Sigma^{-1}$ is the inverse of the covariance matrix. Therefore, the Mahalanobis distance expresses the distance between a data point and a Gaussian distribution in terms of standard deviations after that the distribution has been normalised to unit spherical covariance. 

\begin{figure}
\centering
\includegraphics[width=\hsize]{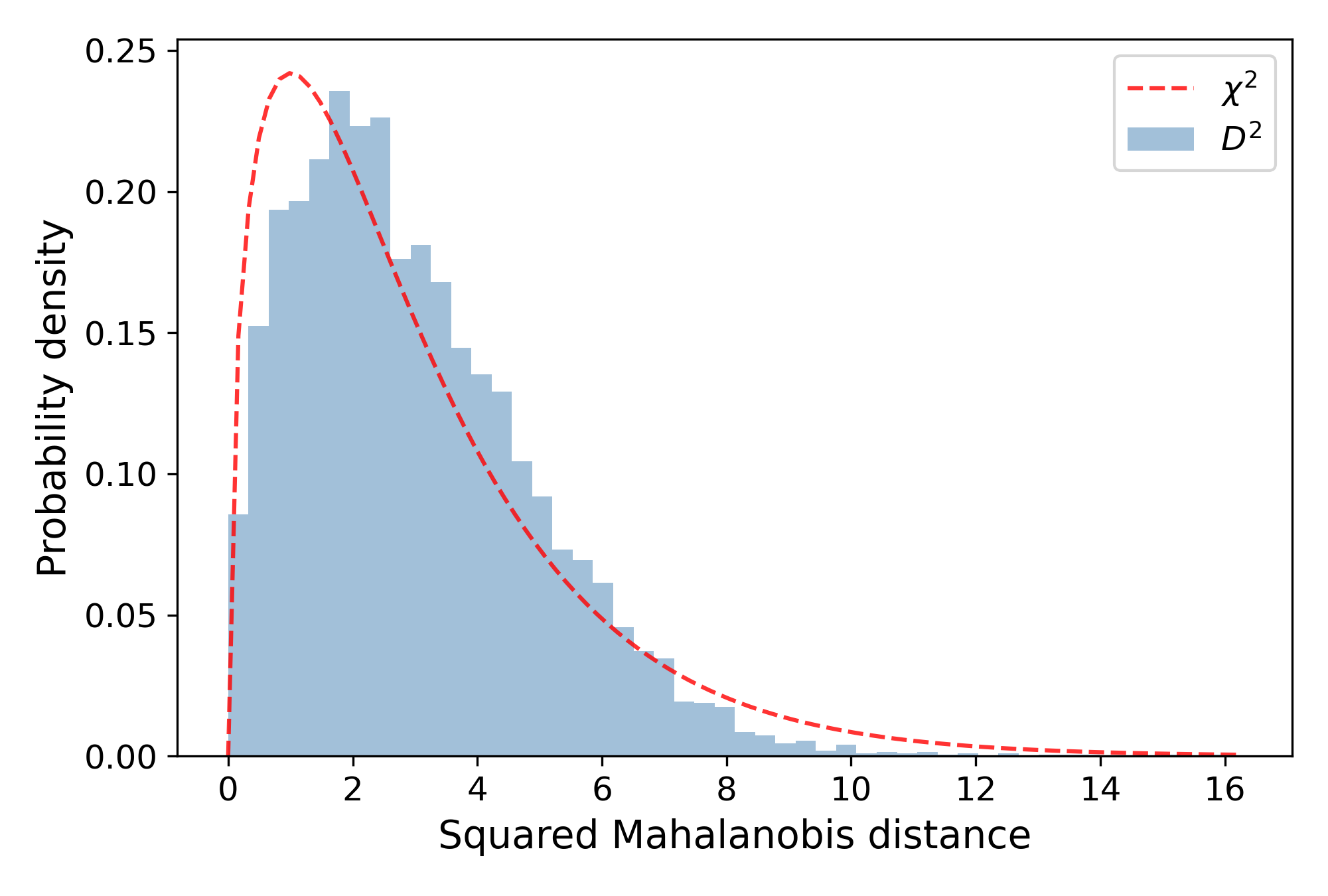}
  \caption{Distribution of squared Mahalanobis distances of the stars assigned to a cluster vs. the chi-square distribution with three degrees of freedom.}
  \label{fig:d2_vs_chi2}
\end{figure}

The theoretical distribution of squared Mahalanobis distances of an $n$-dimensional Gaussian is known: it is the $\chi^2$ distribution with $n$ degrees of freedom. Fig.~\ref{fig:d2_vs_chi2} shows the distribution of Mahalanobis distances for the stars assigned to clusters by our procedure, compared to the $\chi^2$ with $n=3$ degrees of freedom. This figure demonstrates that the distribution of the members of the clusters is somewhat similar to a multivariate Gaussian, but that this is slightly more peaked in comparison to what is seen for the cluster members.

We can therefore use the Mahalanobis distance of each star to refine the detected clusters, select core members, and identify additional plausible members. To establish a meaningful value of the Mahalanobis distance $D_{\rm cut}$, we proceeded as follows. We investigated the internal properties of the clusters as defined by the algorithm and by considering only the stars within different values of $D_{\rm cut}$ (corresponding to the 50th$^{}$, 65th$^{}$, 80th$^{}$, 90th$^{}$, 95th$^{}$, and 99th$^{}$ percentiles of the distribution of original members, see Fig.~\ref{fig:d2_vs_chi2}). Specifically, we checked the distribution of stars in the colour-absolute magnitude diagrams and the metallicity distribution functions for each cluster. A too restrictive cut (50th$^{}$, 65th$^{}$ percentiles) reduces the number of stars in a way that often limits the characterisation of the cluster properties. A too loose cut (95th$^{}$, 99th$^{}$ percentiles), although greatly increasing the number of stars in clusters, leads to a noticeable amount of contamination, hence affecting the cluster properties. As a compromise between a manageable amount of contamination and an increase in purity of the clusters as well as in the number of stars, we decided to use the 80th$^{}$ percentile cut, corresponding to a Mahalanobis distance of $D_{\rm cut} \sim$ 2.13.

Out of the stars in our halo set that did not yet receive a label, we can associate $2104$ additional stars with one of the clusters on the basis of the star being within a Mahalanobis distance of $2.13$ to its closest cluster. There are $1186$ original members falling outside of the Mahalanobis distance cut. In total, $7127$ stars in the halo set $(13.8\%)$ lie within $D_{\rm cut} = 2.13$ of some cluster.

The histogram of the number of cluster members contained within $D_{\rm cut}$ is displayed in Fig.~\ref{fig:n_members_hist} as the dashed red line. The largest cluster, cluster $3,$ has $3032$ such members, while the smallest cluster is cluster $34,$ with $9$ such members. The mean and median sizes of the clusters after the identification of additional members within $D_{\rm cut}$ become $106$ and $49,$ respectively. 

Table~\ref{table:cluster_description} lists the clusters we identified, their statistical significance ($\sigma_i$), the number of members indicated by the clustering ($N_{\rm orig}$) and with additional members ($N_{\rm tot}$), the centroid ($\mu$), and entries of covariance matrix ($\sigma_{ij}$).
To transform the values listed in the table into the corresponding physical quantities, we can use for the means that
\begin{equation*}
    \langle \mu_i \rangle = \frac{2}{\Delta_i} (\langle I_{i}\rangle-I_{i,min}) -1,
\end{equation*}
where $\Delta_i = I_{i,max}-I_{i,min}$, with $i=0..2$ and $I_{i}$ corresponding to $E$, $L_\perp$, and $L_z$, with the minimum and maximum values given in Sec.~\ref{subsec:iom_clustering}. For the covariance matrix, these definitions lead to
\begin{equation*}
    \sigma_{ij} = \frac{4}{\Delta_i \Delta_j}\Sigma_{I_{i,j,}}
\end{equation*}
where $\Sigma_{I_{i,j}}$ is the covariance matrix in integrals-of-motion space.
Table~\ref{table:catalogue} lists the kinematic and dynamical properties for the stars, as well as the cluster with which they were associated by the single-linkage algorithm and that with which they are associated on the basis of their Mahalanobis distance, which is also listed to help assess membership probability. Both tables are available online in their entirety. 

\begin{table*}
\caption{Overview of the characteristics of the extracted clusters. We list their significance, the number of members as determined by the single-linkage ($N_{\rm orig}$) and in total after considering members within a Mahalanobis distance of $2.13$ ($N_{D_{\rm cut}}$). $\mu$ is the centroid of the scaled clustering features as determined by the original members, and $\sigma_{ij}$ represents the corresponding entries in the covariance matrix. The indices 0-2 correspond to the order $E, L_\perp$ , and $L_z$. The full table is made available online.}
\label{table:cluster_description}
\centering
\begin{tabular}{r r r r r r r r}
\hline \hline
                        Label & signif. & $N_{\rm orig}$ & $N_{D_{\rm cut}}$ & $\mu_0$ [$10^{-3}$] & $\mu_1$ [$10^{-3}$] & $\mu_2$ [$10^{-3}$] \\
                        \hline
                        1 & 6.0 & 42 & 33 & -941.5 & -989 & 22 \\
                        2 & 3.1 & 15 & 12 & -640.1 & -952 & 114 \\
                        3 & 6.3 & 2137 & 3032 & -759.6 & -851 & 24 \\
                        4 & 3.3 & 18 & 14 & -607.4 & -450 & 152 \\
                        5 & 4.1 & 69 & 65 & -519.7 & -703 & -38 \\
                        6 & 3.0 & 118 & 122 & -546.2 & -809 & -47 \\
                        7 & 3.8 & 88 & 91 & -533.0 & -914 & -60 \\
                        8 & 3.1 & 60 & 70 & -699.6 & -757 & -201 \\
                        9 & 3.1 & 29 & 17 & -488.2 & -659 & -51 \\
                        10 & 3.1 & 48 & 58 & -505.8 & -869 & -83 \\
                        
                        \hline \hline
                        & 
                        $\sigma_{00}$ [$10^{-6}$] & $\sigma_{01}$[$10^{-6}$]  & $\sigma_{02}$ [$10^{-6}$] & $\sigma_{11}$ [$10^{-6}$] & $\sigma_{12}$ [$10^{-6}$] & $\sigma_{22}$ [$10^{-6}$] \\
                        \hline
                        & 50.7 & 6.3 & 52.7 & 21.0 & 9.2 & 76.9 \\
                        & 57.6 & 37.8 & 30.7 & 28.5 & 19.1 & 25.8 \\
                        & 1324.0 & 255.9 & 23.2 & 3916.6 & 210.6 & 369.2  \\
                        & 42.9 & -11.3 & 44.0 & 49.1 & -19.2 & 81.3 \\
                        & 276.8 & 157.7 & -63.5 & 221.7 & -63.7 & 179.5 \\
                        & 433.2 & -252.2 & 102.0 & 484.5 & -133.5 & 311.3 \\
                        & 136.6 & -49.0 & -51.0 & 410.1 & 231.0 & 418.6 \\
                        & 179.0 & -139.4 & -79.4 & 484.1 & 168.0 & 296.1 \\
                        & 63.4 & 24.2 & -37.9 & 160.9 & -76.5 & 214.2 \\
                        & 648.9 & 31.8 & 216.3 & 241.8 & -65.5 & 162.4 \\
                        
                \end{tabular}
\end{table*}

\begin{table*}
\caption{Overview of fields in our final star catalogue. It contains \textit{Gaia} source ids, heliocentric Cartesian coordinates, heliocentric Cartesian velocities, and the three features we used for clustering. The column significance indicates the significance of the cluster to which a star belongs (according to the original single-linkage assignment), Label$_{orig}$ is the label of the cluster according to the single-linkage process, and Label$_{D_{cut}}$ is the cluster label after selecting the cluster with the smallest Mahalanobis distance $D$ (requiring at least $D<2.13$). The full table is made available online.}
\label{table:catalogue}
\centering
\begin{tabular}{c c c c c c c}
\hline\hline
        source\_id & $x$ & $y$ & $z$ & $v_x$ & $v_y$  & $v_z$ \\
         &  &  &  & [km/s] & [km/s] & [km/s] \\
        \hline
        494990686553088 & -0.82 & 0.07 & -0.92 & -252.3 & -249.5 & 29.7 \\
        2131751183780736 & -0.52 & 0.07 & -0.54 & -262.9 & -247.3 & 45.4 \\
        2412092288843648 & -0.12 & 0.01 & -0.12 & -105.8 & -210.9 & -51.9 \\
        4032119593003520 & -1.56 & 0.17 & -1.49 & 220.9 & -239.3 & 72.0 \\
        4564455019620096 & -0.58 & 0.09 & -0.66 & -204.4 & -83.7 & 57.1 \\
        4641558272441088 & -0.65 & 0.11 & -0.74 & 13.8 & -91.5 & -283.0 \\
        5201244050756096 & -0.76 & 0.12 & -0.81 & 127.5 & -120.1 & -201.5 \\
        5366583111858688 & -0.41 & 0.07 & -0.45 & 169.9 & -108.1 & -93.3 \\
        6932768706211328 & -1.58 & 0.22 & -1.55 & 176.6 & -253.7 & 99.5 \\
        7299795136273664 & -0.74 & 0.13 & -0.72 & 273.0 & -256.0 & 95.1 \\
\hline \hline 
    $E$ & $L_\perp$  & $L_z$  & significance & Label$_{\rm orig}$ & Label$_{D_{\rm cut}}$ & $D$ \\
    $[10^5$ km$^2$/s$^2]$ & $[10^3$ kpc km/s$]$ & [$10^3$ kpc km/s] &  &  &  & \\
        \hline
    -1.192 & 0.56 & -0.06 & 8.9 & 24 & 24 & 1.43 \\
        -1.184 & 0.60 & -0.04 & 8.9 & 24 & 24 & 1.16 \\
        -1.485 & 0.36 & 0.28 & 6.3 & 3 & 3 & 1.46 \\
        -1.141e & 0.43 & 0.10 & 5.5 & 26 & 0 & 2.27 \\
        -1.172 & 0.70 & 1.40 & 3.4 & 30 & 30 & 1.34 \\
        -1.002 & 2.46 & 1.36 & 9.8 & 60 & 60 & 1.79 \\
        -1.133 & 1.85 & 1.14 & 3.6 & 61 & 61 & 1.76 \\
        -1.228 & 0.83 & 1.19 & 3.1 & 38 & 38 & 1.61 \\
        -1.206 & 0.75 & -0.04 & 8.9 & 24 & 24 & 0.83 \\
        -1.042 & 0.71 & -0.06 & 3.6 & 51 & 51 & 1.63\\
\end{tabular}
\end{table*}

\section{Discussion}\label{sec:discussion}

\begin{figure*}[]
    \resizebox{\hsize}{!}
        {\includegraphics[width=\textwidth]{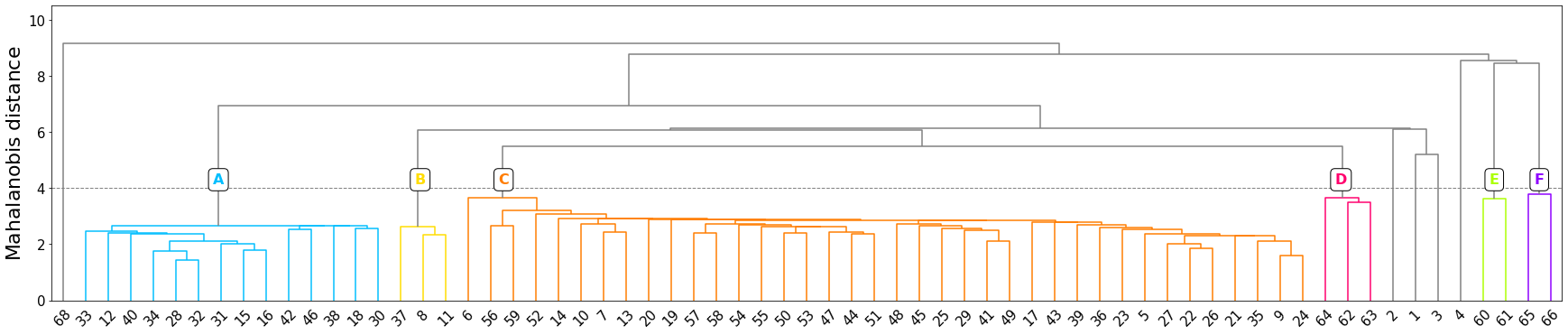}}
        \caption{Relation between the significant clusters according to the single-linkage algorithm, obtained using the Mahalanobis distance in clustering space.}
    \label{fig:dendrogram}
\end{figure*}

The relatively large number of clusters we identified means that  it is important to understand how they relate to each other. We explored their internal hierarchy in our clustering space using the Mahalanobis distance between two distributions,
\begin{equation}
    D^{\prime} = \sqrt{({\boldsymbol \mu_1}-{\boldsymbol \mu_2})^T (\Sigma_1+\Sigma_2)^{-1} ({\boldsymbol \mu_1}-{\boldsymbol \mu_2})}\label{eq:mah_distance_distributions}
,\end{equation}
where $\boldsymbol{\mu}_1, \boldsymbol{\mu}_2$ and $\Sigma_1, \Sigma_2$ describe the means and covariance matrices of the two cluster distributions, respectively, and $D^{\prime}$ gives a relative measure of their degree of overlap. A dendrogram of the clusters according to single linkage by Mahalanobis distance is shown in Fig.~\ref{fig:dendrogram}.

\begin{figure*}[]
\centering
        \includegraphics[width=0.98\textwidth]{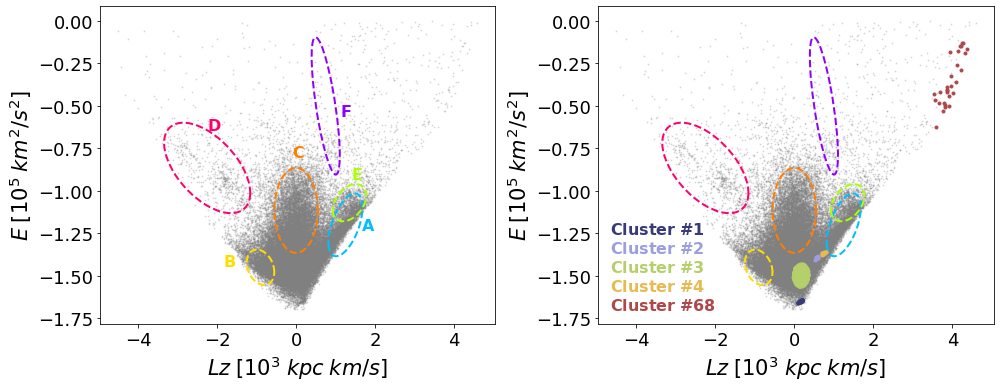}
        \caption{Location in the $E-L_z$ plane of the six main groups identified in Fig.~\ref{fig:dendrogram} together with various individual, more isolated clusters. {\it Left:} Each ellipse shows the approximate locus of the six tentative main groups. {\it Right:} Location of independent clusters (labelled 1, 2, 3, 4, and 68). Each substructure is colour-coded as in Fig.~\ref{fig:dendrogram}, while the independent clusters are colour-coded arbitrarily. As discussed in Sect.~\ref{subsec:prev_works}, substructure B occupies the region dominated by Thamnos1+2, substructure C that of {\it Gaia}-Enceladus, D corresponds to Sequoia, and E to the Helmi streams. Groups A and E overlap in the $E-L_z$ plane, but have very different $L_{\perp}$, resulting in a large Mahalanobis relative distance, as shown in Fig.~\ref{fig:dendrogram}. The orientation of substructure F is peculiar and may indicate that its constituent clusters 65 and 66 should be treated separately.
        }
    \label{fig:IoM_substructures}
\end{figure*}

\subsection{Number of independent clusters}
\label{subsec:howmany}

An initial inspection of the dendrogram shown in Fig.~\ref{fig:dendrogram} reveals a complex web of relations between the significant clusters we extracted with our single-linkage algorithm. Some of these clusters are linked to others only at large distances (clusters 1, 2, 3, 4, or 68), whereas others are clearly grouped together, sometimes even in a hierarchy of substructures. This suggests that the 67 clusters that our methodology identified as significant are not fully independent of each other. The proper assessment of the independence of the different clusters is presented in paper II, \citealt{2022arXiv220102405R}.

As a first attempt to explore the hierarchy we observe in the dendrogram, we tentatively set a limit at Mahalanobis distance $\sim$~4.0 (horizontal dashed line in Fig.~\ref{fig:dendrogram}). This Mahalanobis distance is large enough such that not all $67$ clusters are considered individually \citep[the lower limit is set by the Helmi streams, see the discussion below  and][]{2022A&A...659A..61D}, but also small enough that at least some substructures reported in the literature can be distinguished \citep[see e.g.][]{koppelman2019multiple, naidu2020evidence}. According to this experimental limit, we may tentatively identify six different large substructures (colour-coded in Fig.~\ref{fig:dendrogram}), as well as a few independent clusters. The distribution of these substructures in the $E-L_z$ plane, together with that of the isolated clusters (numbers 1, 2, 3, 4, and 68), is displayed in Fig.~\ref{fig:IoM_substructures}. This figure shows that some of these groups do indeed correspond to previously identified halo substructures (see also Sec.~\ref{subsec:prev_works}). Interestingly, there are also some examples of clusters with hot thick-disk-like kinematics, such as those associated with substructure A, or clusters 2 to 4.

The rich substructure found by the single linkage algorithm within each of these tentative groups deserves further inspection. For instance, the large group labelled C in Fig.~\ref{fig:dendrogram}, which according to its location in the $E-L_z$ plane could correspond to {\it Gaia}-Enceladus \citep[][]{helmi2018merger}, can be clearly split into several subgroups according to the dendrogram. In addition, there are some individual clusters such as cluster 6, linked at large Mahalanobis distances, that might be considered as part of {\it Gaia}-Enceladus. Similarly, clusters belonging to the light blue branch (substructure B), which falls in the region in $E-L_z$ occupied by Thamnos \citep[][]{koppelman2019multiple}, also show different clustering levels. Cluster 37 is linked at a larger Mahalanobis distance than clusters 8 and 11. 

To fully assess the tentative division of our halo set into six main groups and to study the significance and independence of the finer structures within the groups would require a detailed analysis of the internal properties of these clusters and substructures, such as their stellar populations, metallicities, and chemical abundances. We defer such an exhaustive analysis to a separate paper \citep{2022arXiv220102405R}. Nonetheless, as a first example, we here focus our attention on the pair of clusters 60 and 61, identified as substructure E, which are likely part of the Helmi streams \citep[][]{helmi1999debris}.

\begin{figure*}[]
    \resizebox{\hsize}{!}
        {\includegraphics[width=\textwidth]{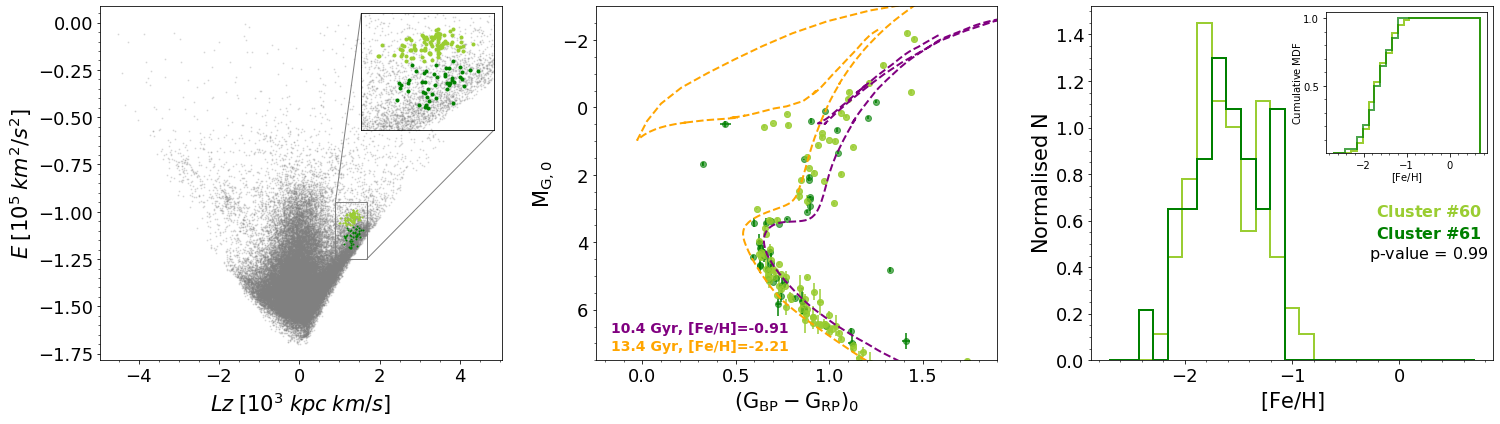}}
        \caption{Characterization of the Helmi streams as detected by the single-linkage algorithm. {\it Left:} Distribution of the 51671 stars analysed in this work as well as clusters 60 and 61 (Helmi streams) in the $E-L_z$ space. {\it Middle:} Colour absolute magnitude diagram for clusters 60 and 61. We overlay isochrones of a 10.4 Gyr ($[$Fe/H$]$=$-0.91$) and a 13.4 Gyr ($[$Fe/H$]$=$-2.21$) old populations from the updated BaSTI stellar evolution models \citep{2018ApJ...856..125H} in the \textit{Gaia} EDR3 photometric system. {\it Right:} Metallicity distribution functions from LAMOST-LRS for clusters 60 and 61, where the cumulative distributions are shown in the inset.}
    \label{fig:helmi}
\end{figure*}

These two clusters are directly linked according to the dendrogram in Fig.~\ref{fig:dendrogram}. The middle and right panels of Fig.~\ref{fig:helmi} show the colour absolute magnitude diagram (CaMD) and the metallicity distribution functions (from the LAMOST-LRS survey) for both clusters. In the CaMD we corrected for reddening using the dust map from \citet[][]{2018A&A...616A.132L} and the recipes to transform into \textit{Gaia} magnitudes given in  \citet[][]{2018A&A...616A..10G}. These figures demonstrate that indeed, the cluster stars depict similar distributions in the CaMD, with ages older than $\sim$~10~Gyr, and metallicities drawn from the same distribution, having a Kolmogorov-Smirnov statistical test p-value of 0.99. All this evidence supports a common origin for both clumps as debris from the Helmi streams. 

The separation of the Helmi streams into two significant clusters in integrals-of-motion space agrees with the recent findings presented in \citet[][]{2022A&A...659A..61D}. In this work, the authors established that the two clumps, which are clearly split in angular momentum space in a similar manner as our clusters 60 and 61, are the result of a resonance in the orbits of some of the stars in the streams. This effect is thus a consequence of the Galactic potential. We preliminarily conclude that analyses such as those presented for the 
Helmi streams here, using the internal properties of the clusters identified by our algorithm, can help us in fully assessing and characterising the different building blocks in the Galactic halo near the Sun.

\subsection{Relation to previously detected substructures}
\label{subsec:prev_works}

\begin{figure*}
\includegraphics[width=\textwidth]{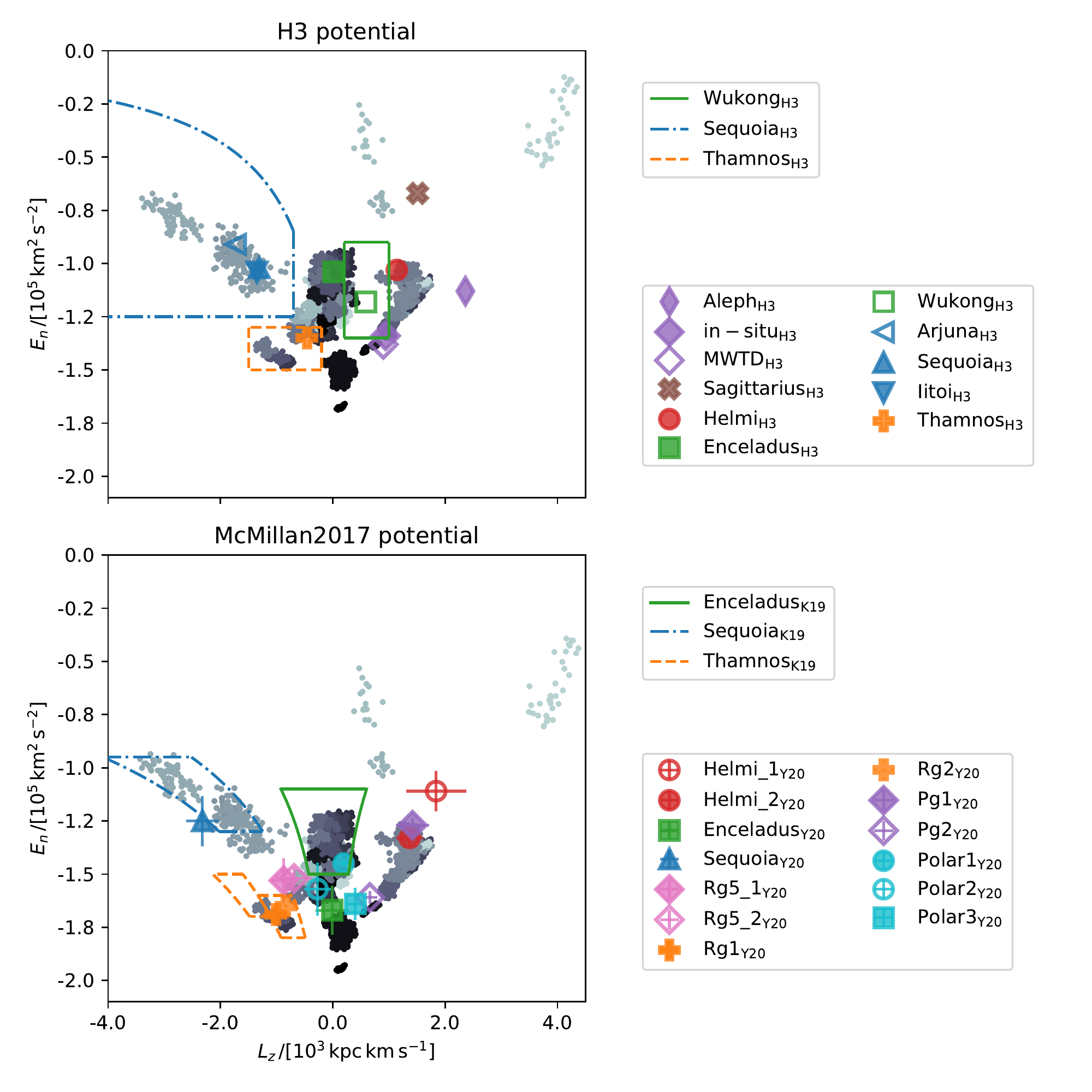}
\caption{Comparison of the clusters identified in the present work with those in the literature.
The $L_z$ and $E$ of the stars in the clusters were re-calculated for consistent comparisons (see text for details).
The upper panel provides a comparison with the results from the H3 survey \citep{naidu2020evidence}, and the bottom panel presents comparisons with selection boundaries from \citet{koppelman2019multiple} and the average kinematic properties of the substructures from \citet{2020ApJ...898L..37Y}.
}
\label{fig:substructure_comparison}
\end{figure*}

Several attempts have been made to identify kinematic substructures in the Milky Way halo after the \textit{Gaia} data releases. 
Here we focus on three studies that have selected or identified multiple kinematic substructures using the data from \textit{Gaia} DR2 \citep{koppelman2019multiple,naidu2020evidence,Yuan2020ApJ...891...39Y}.
Fig.~\ref{fig:substructure_comparison} presents the comparisons in the $L_z-E$ space.
Although the three comparison studies defined kinematic substructures using $L_z$ and $E$ and/or provided average $L_z$ and $E$ of the substructures, they adopted different Milky Way potentials (and a slightly different position and velocity for the Sun). 
We therefore recomputed $L_z$ and $E$ of the stars in our study in the default Milky Way potential of the python package \texttt{gala} \citep{2017JOSS....2..388P}, which was used in \citet{naidu2020evidence}, and in the Milky Way potential of \citet{mcmillan2016mass}, which was used in the other two studies. Reassuringly,  Fig.~\ref{fig:substructure_comparison} shows that the clusters we identified remain tight even in these other Galactic potentials.
In addition to these three studies, we also made a comparison with \citep{Myeong2019MNRAS}, although we note that their identification of substructure was made in subspaces different than in our study.

Based on the data from the H3 survey \citep[][]{2019ApJ...883..107C}, \citet{naidu2020evidence} investigated the properties of a number of substructures by defining selection boundaries simultaneously in dynamical and in the $\alpha$-Fe plane. Their sample consists of stars with a heliocentric distance larger than $3\,\mathrm{kpc}$, and therefore it is complementary to our study, which focuses on stars within 2.5~kpc from the Sun. Even though the spatial coverage is different, the agreement with our work is generally good; we find significant clusters with $L_z$ and $E$ close to the average properties of some structures from \citet{naidu2020evidence}.
However, we also find some differences for Aleph, Wukong, Thamnos, and Arjuna, Sequoia, and I'itoi.

In particular, we do not find significant clusters that would correspond to their Aleph and Wukong. Although several clusters (clusters 12, 27, 39, 52, and 54) have some stars in the Wukong selection box, they are located at the edge. We do not find any clusters that have similar values of $L_z$ and $E$ as the average values of Wukong stars from \citet{naidu2020evidence}. 
The absence of Aleph in our data is likely due to our $|\mathbf{V}-\mathbf{V}_{\mathrm{LSR}}|>210$~km/s selection of halo stars, which would remove essentially all the stars on Aleph-like orbits.
On the other hand, the reason for the absence of Wukong is less clear.

Independently of \citet{naidu2020evidence}, \citet{2020ApJ...898L..37Y} identified a stellar stream called LMS-1 from the analysis of K giants from LAMOST DR6, which they associated with two globular clusters that \citet{naidu2020evidence} associated with Wukong.
Because the findings by \citet{2020ApJ...898L..37Y} 
are based on stars with heliocentric distance of $\sim 20\,\mathrm{kpc}$ and because stars in \citet{naidu2020evidence} are also more distant than our sample, we may tentatively conclude that LMS-1/Wukong might be only prominent at larger distances than 2.5~kpc from the Sun (the distance limit of our sample).
It therefore remains to be seen if the algorithm presented in this work confirms the existence of LMS-1/Wukong when it is applied to a sample that covers a larger volume. 
We note, however, that clusters with $2\sigma$ significance exist in the region of Wukong according to our single-linkage algorithm (see Figure~\ref{fig:significance_map_2sigma}).

We also find some differences when we compare the distribution of our significant clusters with Thamnos and Arjuna, Sequoia, and I'itoi as defined by \citet{naidu2020evidence}.
Our analysis has identified a group of clusters with retrograde motion and with low-$E$ that is clearly separated from {\it Gaia}-Enceladus, which would likely correspond to Thamnos according to the definition of \citet{koppelman2019multiple}, as discussed in the next paragraph. On the other hand, the average $L_z$ and $E$ of Thamnos by \citet{naidu2020evidence} are clearly different, occupying the region where we would probably associate clusters with {\it Gaia}-Enceladus, as shown in the top panel of Fig.~\ref{fig:substructure_comparison}. In the region occupied by Arjuna, Sequoia, and I'itoi, which according to \citet[][]{naidu2020evidence} contains three distinct structures with similar average kinematic properties but different characteristic metallicities, we also find significant clusters with a large retrograde motion and with high orbital energy. Two of our significant clusters (clusters 62 and 63) that have similar $L_z$ and $E_n$ as Arjuna, Sequoia, and I'itoi, while another cluster (64) clearly has different $L_z$ and $E_n$, but
also satisfies a more generous $L_z-E_n$ selection criterion for retrograde structures by \citet{naidu2020evidence}. 
We investigate the relation between all these clusters in detail in \citet{2022arXiv220102405R}.

We now compare our clusters with the selections of {\it Gaia}-Enceladus, Sequoia, and Thamnos from \citet{koppelman2019multiple}. 
We find a number of significant clusters in the three regions identified by these authors to be associated with these objects. 
The analysis presented here (see the bottom panel of Fig.~\ref{fig:substructure_comparison}) suggests that Sequoia might extend towards lower $E$ and that the {\it Gaia}-Enceladus selection might also need to be shifted to lower $E$.
It also suggests that Thamnos stars are more tightly clustered, although still in two significant clumps (see \citet{2022arXiv220102405R}). 
In particular, the region originally occupied by the more retrograde Thamnos component \citep[Thamnos 1 according to][]{koppelman2019multiple} appears fairly devoid of stars. This is partly due to the improvement in astrometry from \textit{Gaia} DR2 to \textit{Gaia} EDR3, especially the reduction in the zero-point offset from $-0.054\,\mathrm{mas}$ \citep{schonrich2019dr2offset} to $-0.017\,\mathrm{mas}$ \citep{2021A&A...649A...1G}.
Thamnos stars originally identified by \citet{koppelman2019multiple} appeared to have lower $L_z$ and higher $E_n$ due to the underestimated parallaxes in DR2.

\citet{Yuan2020ApJ...891...39Y} identified dynamically tagged groups (DTGs) by applying a neural-network-based algorithm, \texttt{StarGO} \citep[][]{2018ApJ...863...26Y}, to a sample of very metal-poor stars ($[\mathrm{Fe/H}]<-2$) within $5\,\mathrm{kpc}$ from the Sun from LAMOST DR3 \citep[][]{2018ApJS..238...16L}.
This implies that their results are more sensitive to the existence of kinematic substructures with low average metallicity.
Here we compare our analysis with their DTGs that they associate with established structures ({\it Gaia}-Enceladus, the Helmi streams, {\it Gaia}-Enceladus, and Sequoia) and Rg5 from \citet{myeong2018discovery}, and DTGs that they classified into several groups depending on their kinematic properties (Rg1, Rg2, Pg1, Pg2, Polar1, Polar2, and Polar3). \citet{Yuan2020ApJ...891...39Y} did not provide separate names for each component of retrograde, prograde, or polar groups. We added a number to distinguish them in Fig.~\ref{fig:substructure_comparison} and to compare them more straightforwardly to our own clusters.
We find significant clusters that have nearly the same $L_z$ and $E$ as many of the groups these authors have identified. There are small offsets between the locations of our clusters and those of the other group they associated with the Helmi streams, their Sequoia, and {\it Gaia}-Enceladus, however.
These offsets might be due to the combined effects of differences in the sample, especially in metallicity and heliocentric distance, and in the method for cluster identification.

\citet{Myeong2019MNRAS} were the first to identify the strongly retrograde substructure Sequoia. Because they defined it in (a normalised) action space, did not characterise its properties in $L_\perp$ and $E$, or published the member stars, we cannot perform a direct comparison  in Figure~\ref{fig:substructure_comparison}.
Nonetheless, \citet{Myeong2019MNRAS} mentioned the presence of a clear peak in the $L_z$ distribution at $L_z\sim -3000\,\mathrm{kpc\,km\,s^{-1}}$ when halo stars with high energy were selected ($E>-1.1\times 10^5\,\mathrm{km^2\,s^{-2}}$ in the \citet{mcmillan2016mass} potential).
This peak would correspond to the most retrograde cluster we identified. We note, however, that if the most retrograde stars are selected in the normalised action space as in \citet{Myeong2019MNRAS}, the sample would also include stars with much lower $E$ and smaller $|L_z|$ \citep{Feuillet2021MNRAS} than those in our most retrograde cluster.

\subsection{Structures with lower statistical significance}

\begin{figure}
\centering
\includegraphics[trim=0 0 1cm 0cm, clip,width=\hsize]{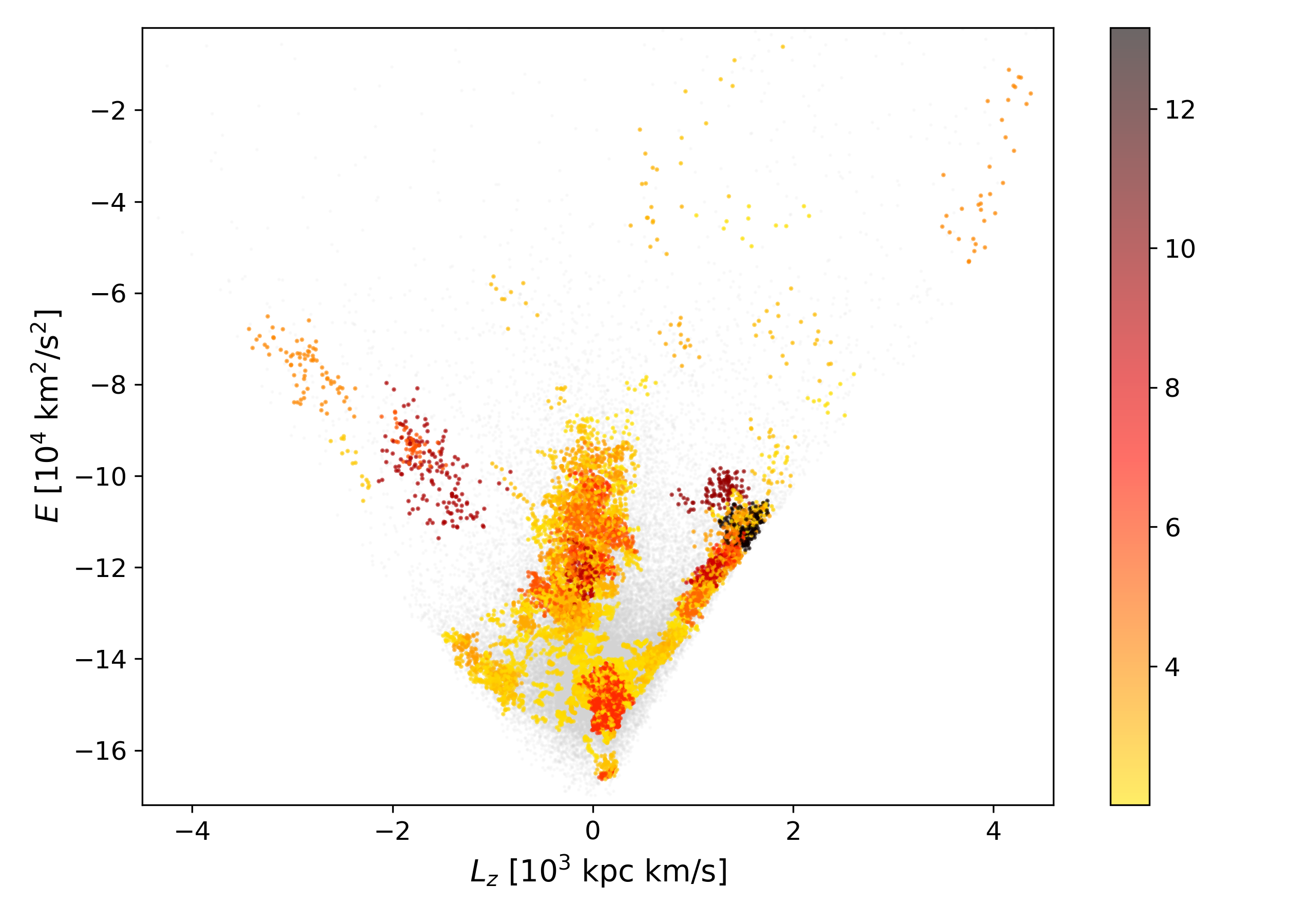}
  \caption{Clusters identified by the single-linkage algorithm with a significance of at least $2\sigma$ . They are colour-coded according to their statistical significance. See for comparison Fig.~\ref{fig:significance_map}.}
  \label{fig:significance_map_2sigma}
\end{figure}

Fig.~\ref{fig:significance_map_2sigma} displays clusters in the data set with a significance level of at least $2\sigma$. We identify $362$ clusters at this level, with $11150$ ($21.6$ \%) original members. In total, $11311$ stars ($21.9$ \%) lie within a Mahalanobis distance of 2.13 to one of the $2\sigma$ significance clusters.

A large number of the stars in the clusters with significance between 2 and 3$\sigma$ are located in the region between {\it Gaia}-Enceladus and the hot thick disk in $E-L_z$ space. This densely populated region makes it harder to identify highly significant overdensities in comparison with our randomised data sets. Thus, it might be argued that the additional clusters of lower significance may be interesting in any case. A possible way to assess a more realistic level of acceptable statistical significance could be to train the algorithm on simulations such that the ratio of extracted signal to undesired false positives can be maximised.

\subsection{Summary}

Our single linkage algorithm has identified $67$ statistically significant clusters in integrals-of-motion space, many of which are rather compact. The discussions of the previous sections suggest that they can be tentatively grouped into six main independent substructures, with some room for further splitting as well as merging \citep[see][for a more thorough analysis]{2022arXiv220102405R}. 

Although we carried our analysis neglecting the effect of errors, we verified their impact on the results in two ways. Firstly, we compared the size of the errors in integrals-of-motion space of these stars to the size of the clusters. To estimate the sizes, we used the determinant of the covariance matrix that describes each cluster and that of the covariance matrix describing the measurement uncertainties of each individual cluster star member. The median ratio of the determinants per cluster is 5.3$\times10^{-4}$ (while it is 3.12$\times 10^{-6}$ for stars from the {\it Gaia} RVS halo sample). This implies that the median volume occupied by the error ellipsoid of a star in integrals-of-motion space compared to that of the cluster it belongs to is 2.3\% (and 0.18\% for the RVS subset). Clusters with a small number of stars tend to be slightly more affected by the uncertainties. 
We also determined how uncertainties may affect the clustering analysis. To this end, we used the {\it Gaia} RVS subset of our halo sample (which has lower uncertainties), and  
convolved the stars values with errors characteristic of the extended sample. We then compared the results obtained from running the algorithm on both the original and the error convolved subsamples. 
The effect of errors is to cause clusters to become inflated, which typically implies that the significance of a cluster is lower than it would be without errors. Although the clusters we identified as significant in this comparison are not always identical, we recovered the majority of the clusters as they occupy the same regions of integrals-of-motion space. As a consequence, the grouping into independent substructures is also robust to the measurement errors.

Our algorithm tends to extract many local overdensitites within what we would associate with a single \added{(}accreted\added{)} object on the basis of the discussions presented in  Secs.~\ref{subsec:howmany} and  \ref{subsec:prev_works}.
On the one hand, this could be due to the true morphology of the \added{(}accreted\added{)} object in integrals-of-motion space, which is expected to be split into substructures corresponding to individual streams crossing a local volume \citep[see e.g.][]{gomez2010identification}. On the other hand, it could be related to the difficulty of obtaining a high statistical significance for objects with a large extent in integrals-of-motion space. For example, in the case of {\it Gaia}-Enceladus, it might be argued that this is due to the steep gradient in the density of stars in the integrals-of-motion space, particularly in energy. The single linkage algorithm generally links stars with higher binding energy {\it \textup{earlier}} in the merging process because there are more such stars. From there it will form connected components that will grow towards lower binding energies. Thus the region occupied by {\it Gaia}-Enceladus is likely to be contaminated by high binding energy stars linked in such a way that the statistical significance of the less bound stars in what could be the {\it Gaia}-Enceladus region is never evaluated on its own. 

A large number of clusters (although smaller than reported here) have also been found by \citet{Yuan2020ApJ...891...39Y} using \textit{Gaia} DR2 data. It is perhaps to be expected that the improvement in the astrometry provided by \textit{Gaia} EDR3 data has led to the discovery of even more substructures. Nonetheless, it is likely that we did not identify all{\it } the individual kinematic streams in our data set even after application of the HDBSCAN algorithm in velocity space on our single-linkage clusters. The velocity dispersions of the $232$ (HDBSCAN) subgroups (see Fig.~\ref{fig:v_space_dispersions}) are somewhat higher than expected  for individual streams \citep{1999MNRAS.307..495H}, although this might partly be due to orbital velocity gradients because of the finite volume considered. Our ability to distinguish these moving groups is probably restricted by the number of stars (see Sec.~\ref{subsec:clustering_vspace}). Furthermore, as discussed earlier, the region of low $E$ and low $L_z$, given its high stellar density, is particularly hard to disentangle. 

\section{Conclusions}\label{sec:conclusions}

We have constructed a sample of halo stars within 2.5 kpc from the Sun using astrometry from \textit{Gaia} EDR3 and radial velocities from \textit{Gaia} DR2 supplemented with data from various ground-based spectroscopic surveys. Our goal was to systematically identify substructures in integrals-of-motion space that could tentatively be associated with merger events. To this end, we applied the single linkage algorithm, which returns a set of connected components or potential clusters in the data set. 
We determined the statistical significance of each of these candidate clusters by comparing the observed density of the cluster with that of an artificially generated halo, obtained by scrambling the velocity components of our real halo set. As a statistical significance threshold, we required that the density within an ellipsoidal contour covering 95\% of the cluster distribution to be more than three standard deviations away from the mean expected density of the artificial data sets. Our final cluster labels were extracted by tracing all (possibly hierarchically overlapping) significant clusters in the merging tree and selecting the nodes at which the statistical significance was maximised. In this way, we identified $67$ statistically significant clusters.

To obtain an indication of how likely it is for a star in our data set to belong to a cluster, we modelled each cluster as a Gaussian probability density. We then determined the Mahalanobis distance between any star and the cluster. As a guidance, a Mahalanobis distance of $D_{\rm cut} \sim 2.13$ roughly contains 80\% of the stars in a cluster, and therefore this value can be used to identify core members that may be interesting for follow-up.
We found that $\sim 13.8\%$ of the stars in our sample can be associated with a significant cluster according to this criterion. 

Our findings are summarised in Table~\ref{table:cluster_description} and Table~\ref{table:catalogue}, which list the characteristics of the substructures and the dynamical properties of the stars together with their Mahalanobis distances. These tables are made available in electronic format upon publication, when we also release the source codes on Github. 

We also identified subgroups in velocity space ($v_R, v_\phi, v_z$) by applying the HDBSCAN algorithm separately on each statistically significant cluster. In this way, we extracted $232$ streams, some of which clearly correspond to subgroups resulting from a stream wrapping around its orbit that is observed as it crosses a local volume.

Furthermore, we also investigated the internal relation between the clusters and how they map to previously established structures. We were tentatively able to group the clusters into roughly six main independent structures, leaving aside a number of independent clusters. Their characterisation and interrelation is the focus of our paper II \citep{2022arXiv220102405R}. In that work, we scrutinise their reality in detail in terms of consistency in stellar populations, chemical abundances, and metallicity distributions, for example. We may conclude that we have made significant steps towards a robust characterisation of the substructure in the halo near the Sun.

\begin{acknowledgements}
We are grateful to the referee for the constructive report.
We have used Python and the following libraries to implement our method in code: Vaex, for efficient handling of the data set and data exploration \citep{Breddels_2018}. Scipy, for implementation of the single linkage algorithm and chi-square distribution \citep{2020SciPy-NMeth}. HDBSCAN for extracting substructure in velocity space \citep{McInnes2017} and NumPy and Matplotlib for utility functions \citep{harris2020array, hunter2007matplotlib}. We gratefully acknowledge financial  support from a Spinoza prize from the Dutch Research Council (NWO) and HHK gratefully acknowledges financial support from the Martin A. and Helen Chooljian Membership at the Institute for Advanced Study.
This work has made use of data from the European Space Agency (ESA) mission {\it Gaia} (\url{https://www.cosmos.esa.int/gaia}), processed by the {\it Gaia} Data Processing and Analysis Consortium (DPAC, \url{https://www.cosmos.esa.int/web/gaia/dpac/consortium}). Funding for the DPAC has been provided by national institutions, in particular the institutions participating in the {\it Gaia} Multilateral Agreement.
This work also made use of the Third Data Release of the GALAH Survey \citep[][]{2021MNRAS.506..150B}. The GALAH Survey is based on data acquired through the Australian Astronomical Observatory. We acknowledge the traditional owners of the land on which the AAT stands, the Gamilaraay people, and pay our respects to elders past and present. 
This paper has made as well use of APOGEE DR16 data part of the SDSS IV scheme. Funding for the Sloan Digital Sky Survey IV has been provided by the Alfred P. Sloan Foundation, the U.S. Department of Energy Office of Science, and the Participating Institutions. 
We have made use of RAVE data for this work, see the RAVE web site at \url{https://www.rave-survey.org}.
Guoshoujing Telescope (the Large Sky Area Multi-Object Fiber Spectroscopic Telescope LAMOST) is a National Major Scientific Project built by the Chinese Academy of Sciences. Funding for the project has been provided by the National Development and Reform Commission. LAMOST is operated and managed by the National Astronomical Observatories, Chinese Academy of Sciences. 

\end{acknowledgements}

\bibliographystyle{aa} 
\bibliography{references.bib}

\end{document}